\documentclass{aa}
\usepackage{txfonts}
\usepackage{graphicx}
\usepackage{caption}
\usepackage{subcaption}
\usepackage{color}

\usepackage{tikz}
\usetikzlibrary{arrows,shapes,backgrounds}

\begin{document}
\titlerunning{}
\authorrunning{Masoura et al.}

\title{The relation between AGN type and host galaxy properties}

\author{V. A. Masoura \inst{1,2}, G. Mountrichas\inst{3,1}, I. Georgantopoulos\inst{1}, M. Plionis\inst{1,2}}
          
    \institute {National Observatory of Athens, V.  Paulou  \& I.  Metaxa, 15 236 Penteli, Greece
              \email{vassi.lia.mas@gmail.com}
           \and
             Section of Astrophysics, Astronomy and Mechanics, Department of Physics, Aristotle University of Thessaloniki, 54 124, Thessaloniki, Greece
             \and
             Instituto de Fisica de Cantabria (CSIC-Universidad de Cantabria), Avenida de los Castros, 39005 Santander, Spain}

\abstract {We use 3,213 Active Galactic Nuclei (AGNs) from the $\it{XMM}$-XXL northern field to investigate the relation of AGN type with host galaxy properties. Applying a Bayesian method, we derive the hardness ratios (HRs)  and through these the hydrogen column density ($\rm N_H$) for each source. We consider as absorbed sources (type-2) those with $\rm N_H > 10^{21.5}\, \rm{cm^{-2}}$. We examine the star formation rate (SFR) as well the stellar mass (M$_*$) distributions for both absorbed and unabsorbed sources. Our work indicates that there is no significant link between the AGN type and these host galaxy properties. Next, we investigate whether the AGN power, as represented by its X-ray luminosity, $\rm L_X$, correlates with any deviation of the host galaxy’s place from the so-called Main Sequence of galaxies, and we examine this separately for the obscured and the unobscured AGN populations. To take into account the effect of M$_*$ and redshift on SFR, we use the normalised SFR (SFR$_{norm}$). We find that the correlation between $\rm L_X$ and SFR$_{norm}$, follows approximately the same trend for both absorbed and unabsorbed sources, a result that favours the standard AGN unification models. Finally, we explore the connection between the obscuration, $\rm N_H$, and the SFR. We find that there is no relation between these, suggesting that the obscuration is not related to the large scale SFR in the galaxy.}

   
\maketitle

\section{Introduction}

Active Galactic Nuclei (AGNs) are among the most luminous radio, optical and X-ray sources in the Universe. They are powered by accretion onto supermassive black holes (SMBHs), M$_{\rm BH}$ $\geqslant$ $10^6$ M$_\odot$, which are located in their centres. Despite the difference in physical scale between the SMBH and the galaxy spheroid \citep[nine orders of magnitude;][]{Hickox2011}, there has been evidence of a causal connection between the growth of the SMBH and the host galaxy evolution. These pieces of evidence come from both observational \citep[e.g.][]{Magorrian1998, Ferrarese2000} and theoretical studies \citep[e.g.][]{Hopkins2006, Hopkins2008, DiMatteo2008}.

One popular method to study the  AGN$-$galaxy coevolution is via examining the correlation, if any, of the SMBH activity and the star formation of the host galaxy, at different epochs \citep[e.g.][]{Rovilos2012, Rosario2013, Chen2013, Hickox2014, Stanley2015, Rodighiero2015, Aird2016, Aird2019}. The first attempts to examine this coevolution were hampered by low number statistics \citep[e.g][]{Lutz2010, Page2012}, field-to-field variations \citep{Harrison2012}. Recent studies have examined the star formation rate (SFR)$-$AGN power relationship using data from wider fields \citep[e.g.][]{Lanzuisi2017, Brown2019}. Some of these studies take into account the evolution of SFR with redshift, by splitting their results into redshift bins. However, most studies do not compare the SFR of AGN with that of similar systems that do not host an AGN. \cite{Masoura2018} used data from both the X-ATLAS and $\it{XMM}$-XXL North fields and found evolution of the SFR of AGN with stellar mass and redshift. The mean SFR of AGN at fixed stellar mass and redshift is higher than star-forming galaxies that do not host an AGN, in particular at $\rm z>1$ \citep[see also][]{Florez2020}.They also examined the SFR$_{norm}$, that is the observed SFR divided by the expected SFR at a given $\rm M_*$ and redshift, as a function of the AGN power \citep[see also][]{Mullaney2015, Bernhard2018, Grimmett2020}. Based on their results, the AGN enhances or quenches the star formation of its host galaxy depending on the location of the host galaxy relative to the star formation main sequence. \cite{Bernhard2018} used data from COSMOS field and found that the SFR$_{norm}$ of powerful AGN has a narrower distribution that is shifted to higher values compared to their lower X-ray luminosity ($\rm L_X$) counterparts. However, the mean SFRs are consistent with a flat relationship between SFR and $\rm L_X$. In addition to the aforementioned effects, it has been pointed out that different observed trends could be the consequence of different binning \citep {Volonteri2015, Volonteri2015a, Lanzuisi2017}. A possible explanation could be the differences, in timescales, between the black hole accretion rate and the SFR. Specifically, AGNs may be expected to vary on a wide range of timescales (hours to Myr) that are extremely short compared to the typical timescale for star formation (100 Myr) \citep{Hickox2014}. Recently, \cite{Grimmett2020} presented a novel technique that removes the need for binning, by applying a hierarchical Bayesian model. Their results confirmed those of \cite{Bernhard2018}, that higher luminosity AGN have a tighter physical connection to the star-forming process than lower X-ray luminosity AGN, in the redshift range probed by their dataset.

Most of the radiation emitted by AGNs is obscured from our view, due to the presence of material between the central source and the observer \citep{Fabian1999, Treister2004}. Obscured AGNs consist up to $\sim$ (70$\%$ $-$ 80$\%$) of the total AGN population \citep[e.g.][]{Akylas2008, Georgakakis2017}. As a result, obscuration presents a significant challenge to reveal the complete AGN population and understand the cosmic evolution of SMBHs. The main reasons of obscuration are, both, the fuelling of the SMBH (inflows of gas) and the AGN feedback \citep[for a review see][]{Hickox2018a}. In terms of the different obscuring material, X-ray energies are obscured by gas, whereas ultraviolet (UV) and optical wavelengths are extincted by dust.

According to the simple unification model, \citep[e.g.][]{Antonucci1993, Urry1995, Netzer2015}, optical and UV emission is produced by the accretion disc around the SMBH. The X-ray emission is produced by the hot corona and is associated with the accretion disk, having a tight relation with the UV and optical emission \citep{Lusso2016}. To account for the strong infrared (IR) emission \citep{Sanders1989, Elvis1994}, the model includes a torus of gas and dust that forms around the central engine (SMBH and accretion disc). Specifically, the dust grains \citep[see][]{Draine2003} of the toroidal structure are heated by the radiation from the central engine, which is then re-emitted at larger wavelengths \citep[IR emission;][]{Barvainis1987}. The orientation of the dusty torus determines the amount of obscuring material along the observer’s line of sight to the central regions. Thus, type 1 refers to the face-on (unobscured) and type 2 to the edge-on (obscured) AGNs. As a result, one of the main predictions of the simple unification model is that type 1 and type 2 AGNs should live in similar environments and thus have similar host galaxy properties.

On the other hand, according to the evolutionary model, different levels of obscuration correspond to different stages of the growth of the SMBH \citep{Ciotti1997, Ciotti2001, DiMatteo2005, Hopkins2006a, Bournaud2007, Gilli2007, Somerville2008, Treister2009, Fanidakis2011a}. The main idea is that the AGN growth coincides with host galaxy activity which is likely to obscure the AGN (type 2 $-$ phase). Eventually, the powerful AGN pushes away the surrounding star forming material ("blowout"). As a result further star formation is halted ("quenching") revealing the unobscured AGN (type 1 $-$ phase). In the context of this model the same galaxy material (dust and gas), that obscures the AGN, may also produce the star formation. It has been claimed that obscuration can occur not only in the regime around the accretion disk, but also in galaxy scale \citep[e.g.][]{Maiolino1995, Malkan1998, Matt2000, Netzer2015, Circosta2019, Malizia2020a}. This large-scale extinction, is typical in models where SMBH$-$galaxy co-evolution is driven by mergers \citep[e.g.][]{Hopkins2008, Alexander2012, Buchner2016}. Early studies indicated large scale morphological differences between type 2 and type 1 AGN, in Seyfert galaxies, with the former having more frequent asymmetric/disturbed morphologies \citep{Maiolino1997}.

Many different approaches are used in the literature to examine the aforementioned theoretical picture of the AGN obscuration, and its possible relation with large scale properties of the host galaxy. Investigating the relation between AGN type and host galaxy properties presents a popular approach to achieve that. The most common way, is to examine whether or not there are differences in various host galaxy properties of obscured and unobscured AGN. \cite{Merloni2014} reported no significant differences between the mean M$_*$ and SFRs. On the other hand, \cite{Zou2019}, claimed that unobscured AGNs tend to have lower M$_*$ than the obscured ones. However, according to the same study, SFRs are similar for both classifications. According to \cite{Chen2015}, obscured AGNs have higher IR star formation luminosities by a factor of approximately 2, than unobscured ones. Furthermore, they studied the connection between obscuration and SFR and based on their results there is an increase of the obscured fraction with SFR. Nevertheless, in their work the examined sample consists of luminous quasars and the sources are classified as obscured and unobscured based on their optical classification. Studies that used X-ray sources found that the correlation of SFR and X-ray absorption is either non-existent or mild \citep[e.g.][]{Rovilos2007, Rovilos2012, Rosario2012, Stemo2020}. Therefore, it is still an open question, whether there is a connection between X-ray obscuration and host galaxy properties.

In this work, we use X-ray AGN in the {\it{XMM}}-XXL field and examine the relation, if any, between X-ray absorption and galaxy properties. Applying a Bayesian method, we derive the hardness ratios (HRs) for the examined sample and through them the hydrogen column density ($\rm N_H$). Sources with $\rm N_H > 10^{21.5}\, \rm{cm^{-2}}$ considered as absorbed. Taking into account the $\rm N_H$, we examine the SFR as well the M$_*$ distributions for both populations (absorbed and unabsorbed) and we re-examine the connection between the $\rm L_X$ and the SFR of the host galaxy. Additionally, we explore the connection between the $\rm N_H$ with the SFR. We assume a $\Lambda$CDM cosmology with $\Omega_m$ = 0.3, $\Omega_{\Lambda}$ = 0.7 and $\rm H_0$ = 70 km s$^{-1}$Mpc$^{-1}$.

\section{Data}

\subsection{The $\it{XMM}$-$\rm XXL$ survey}

To carry out our analysis we use sources from the $\it{XMM}$-XXL field. The $\it{XMM-Newton}$ XXL survey \citep[$\it{XMM}$-XXL][]{Pierre2016} is a medium-depth X-ray survey that covers a total area of 50 $\deg^2$ split into two fields equal in size, the $\it{XMM}$-XXL North (XXL-N) and the $\it{XMM}$-XXL South (XXL-S). The data, used for the current work, come from the equatorial sub-region of the XXL-N, which consists of 8,445 X-ray sources. Of these X-ray sources, 5,294 have SDSS counterparts and 2,512 have reliable spectroscopy \citep{Menzel2016, Liu2016}. The data reduction, source detection and sensitivity map construction follow the methods described by \cite{Georgakakis2011a}.

\subsection{The X-ray AGN sample}
The XXL sample used in our analysis is the same sample used in \cite{Masoura2018}. The details on source selection and SED fitting analysis, are provided in their Sections 2 and 3, respectively (see also their Tables 1 and 2). Here, we outline the most important parts. 

The dataset consists of 3,213 X-ray selected AGN from the XXL-N field, within a redshift range of $\rm 0.03<z<3.5$. 1849 sources have spectroscopic redshift \citep{Menzel2016} and 1364 have photometric redshift (photoz). Photoz were estimated using TPZ, a machine-learning algorithm \citep{Kind2013}, following the process described in \cite{Mountrichas2017} and \cite{Ruiz2018}. 

All our sources have available optical photometry from SDSS. mid-IR ({\it{WISE}}) and near-IR (VISTA) photometry was obtained following the likelihood ratio method \citep{Sutherland_and_Saunders1992}. We use catalogues produced by the HELP\footnote{The {\it Herschel} Extragalactic Legacy Project (HELP, http://herschel.sussex.ac.uk/) is a European funded project to analyse all the cosmological fields observed with the {\it Herschel} satellite. HELP data products are accessible on HeDaM (http://hedam.lam.fr/HELP/)} collaboration to add far-IR counterparts. HELP provides homogeneous and calibrated multiwavelength data over the {\it Herschel}  Multitiered Extragalactic Survey \citep[HerMES,][]{Oliver2012a}  and the  H-ATLAS survey (Eales et al. 2010). The  strategy adopted by HELP is to build a master list catalog of objects for each field (Shirley et al. 2019) and to use the near-IR sources from IRAC surveys as prior information for the IR maps. The  XID+ tool \citep{Hurley2017}, developed for this purpose, uses a  Bayesian probabilistic framework  and works with prior positions. At the end, a flux is  measured for all the near-IR sources of the master list. In this work, only SPIRE fluxes are considered, given the much lower sensitivity of the PACS observations for this field \citep{Oliver2012}. 1,276 X-ray AGN have {\it Herschel} photometry.

In our analysis, we make SFR estimations, through SED fitting. For this reason, we require our sources to have at least, $\it{WISE}$ (W1-W4) or $\it{HERSCHEL}$ detection, in addition to optical photometry.  SFR estimations for sources without $\it{HERSCHEL}$ photometry have been calibrated using the relation presented in \cite{Masoura2018}(for more details see their section 3.2.2 and Fig. 4).

\begin{figure}
\centering
  \includegraphics[width=0.99\linewidth]{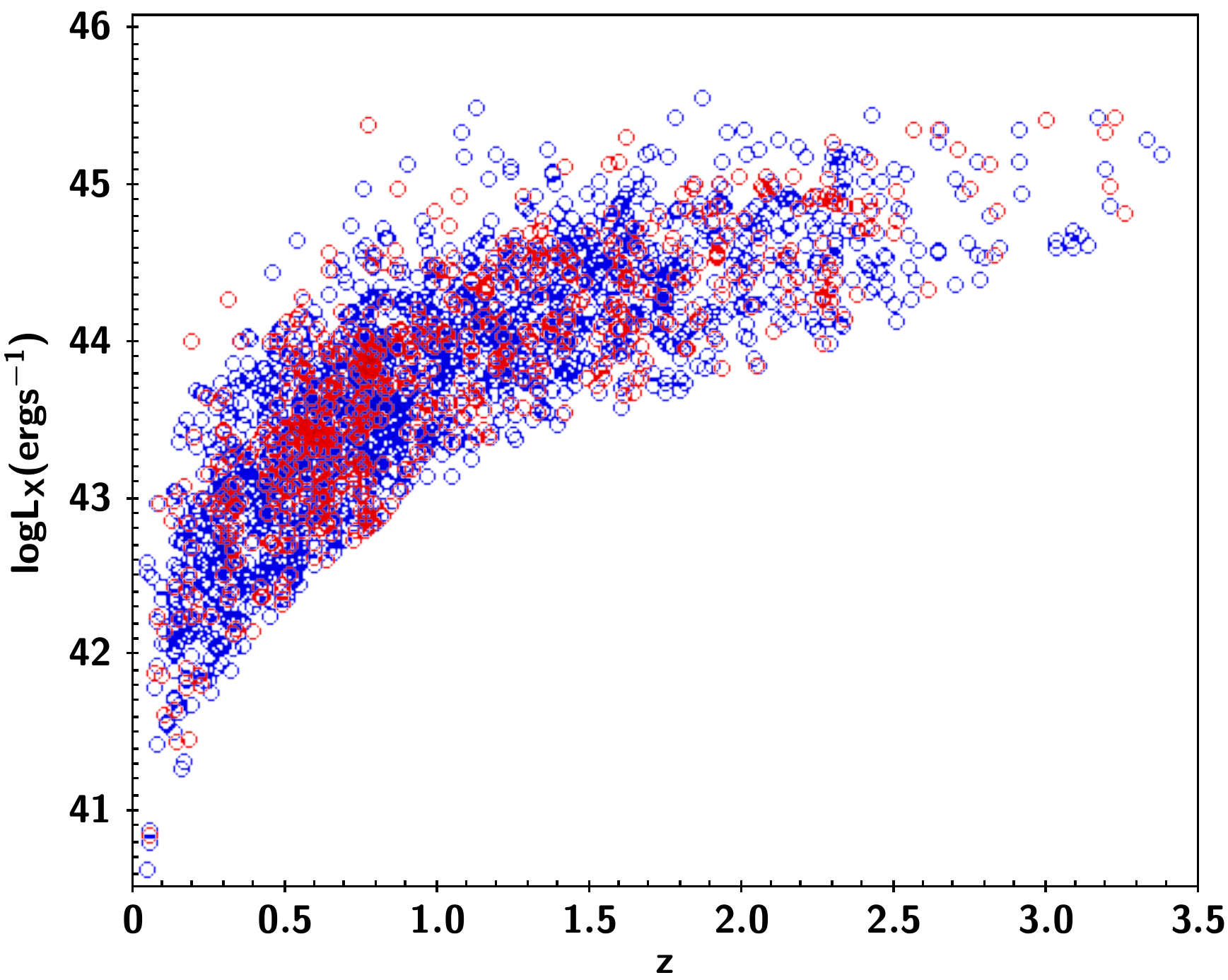}
      \caption{The observed X-ray luminosity as a function of redshift. Blue and red colours refer to the unabsorbed and absorbed sources, respectively. There is a selection bias against low luminosity sources (log $\rm L_X$ < 43 erg s$^{-1}$) at high redshifts (z > 1). The effect of this incompleteness is taken into account (for details see section \ref {sec_sfrnorm_lx}).}
\label{Lx_z_1}
\end{figure}

In table \ref{sample}, we present the number of sources based on the available photometry, spectroscopy and X-ray absorption. The observed $\rm L_X$ of our sample, are estimated in the hard energy band (2$-$10 keV). Our X-ray sources are selected to have $\rm L_X$ (2$-$10 keV) > $10^{41}$ erg s$^{-1}$, which minimises contamination from inactive galaxies. The observed X-ray luminosity as a function of redshift is presented in Fig. \ref{Lx_z_1}. Based on the latter plot, there is a selection bias against low luminosity sources (log $\rm L_X$ < 43 erg s$^{-1}$) at high redshifts ($\rm z > 1$). To account for this effect we adopt the $\rm \frac {1}{V_{max}}$ method \citep[e.g.][see section \ref{sec_sfrnorm_lx}]{Schmidt1968, Akylas2006}.

\begin{table*}
\caption{Number of spectroscopic and photometric sources with SDSS, $\it{WISE}$ or $\it{HERSCHEL}$ photometry in our sample. The number of AGNs used in this study appear in bold. The sources are classified as absorbed or unabsorbed based on their $\rm N_H$. We consider as absorbed those sources with $\rm N_H > 10^{21.5}\, \rm{cm^{-2}}$. The numbers in parentheses refer to the absorbed AGNs.}

\centering
\setlength{\tabcolsep}{2.3mm}
\begin{tabular}{cccc}
      \hline
X-ray selected AGN&Total number&Specz&Photoz\\
       \hline
SDSS&5294&2512&2782\\
\hline
SDSS + $\it{WISE}$ or $\it{HERSCHEL}$&\textbf{3213}\,(\textbf{808})&\textbf{1849}\,(\textbf{454})&\textbf{1364}\,(\textbf{354})\\
\hline
\label{sample}
\end{tabular}
\end{table*}

\section{Analysis}
In this section, we describe the steps we have followed in our analysis. The AGN host galaxy properties have been estimated through SED fitting (section \ref{prop}). Section \ref{X} describes the estimation of the X-ray properties of our sample and the methodology we have followed to account for selection biases.

\subsection{Host galaxy properties}
\label{prop}
SFR and M$_*$ were estimated using the CIGALE code version 0.12 \citep[Code Investigating GALaxy Emission][]{Burgarella2005, Noll2009, Boquien2019}. The emission associated with AGN is modelled using the \cite{Fritz2006} library, as described in \cite{Ciesla2015}. This allows us to disentangle the AGN IR emission from the IR emission of the host galaxy and derive more accurate SFR measurements. Sources with reduced $\chi ^2>5$ have been excluded from our analysis. For more details on the SED fitting, the models used and their parameter values, see \cite{Masoura2018}. 

\cite{Masoura2018} (see their Fig. 7), found that the average SFR of AGN host galaxies presents a similar evolution with stellar mass and redhsift with that of star-forming systems \citep[e.g][]{Brinchmann2004, Elbaz2007, Daddi2007, Magdis2010, Salmon2015, Schreiber2015}. To account for the evolution of the MS, we follow their approach \citep[see also e.g., ][]{Bernhard2018} and make use of the SFRs$_{norm}$ parameter to examine whether the AGN type plays any systematic role in deviations (or dispersion) around it. This quantity is equal to the observed SFR, divided by the expected SFR at a given $\rm M_*$ and redshift (SFRs$_{norm}$= SFR/SFR$_{MS}$). To estimate the SFR$_{MS}$ we use equation 9 of \cite{Schreiber2015}.
 
\subsection{X-ray absorption}
\label{X}

To classify the AGN as X-ray absorbed and unabsorbed we need to estimate their hydrogen column density. For that purpose, we first apply a Bayesian method to calculate the HR and then infer the $\rm N_H$ of each source. A detailed description of this process is provided below:
 
\subsubsection{X-ray colours}
HR or X-ray colour, is used to quantify and characterise weak sources and large samples. To alleviate the downsides of the classical definition (e.g. Gaussian assumption for the error propagation of counts for faint sources with a significant background, background subtraction), we apply the Bayesian Estimation of Hardness Ratios code \citep[BEHR;][]{Park2006}. This method applies a Bayesian approach to account for the Poissonian nature of the observations. For the estimations, we use the number of photons in the soft (0.5$-$2.0 keV) and the hard (2.0$-$8.0 keV) bands, provided in \cite{Liu2016} catalogue. The hardness ratio calculations are  based on the following equation:

\begin{equation}
\rm HR = \rm \frac {H-S}{H+S},
\end{equation}
where S and H the counts in the soft and the hard bands, respectively. 

\subsubsection{Hydrogen column density}
\label{abs_section}

\begin{figure}
\centering
  \includegraphics[width=0.99\linewidth]{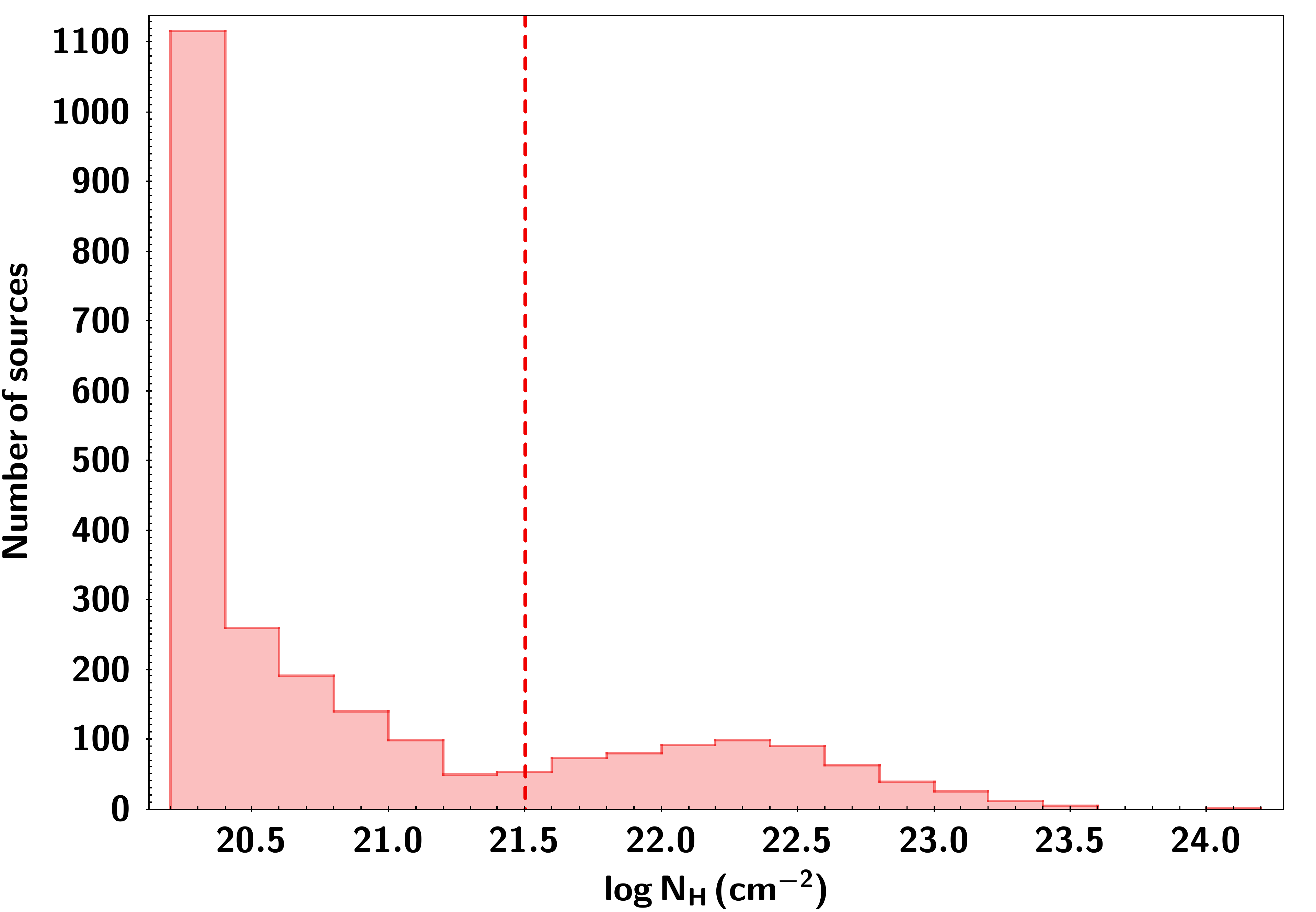}
  \caption{Hydrogen column density, $\rm N_H$, distribution of the AGN sample. The vertical, dashed line presents the $\rm N_H$ cut applied in our analysis to split the sources into absorbed and unabsorbed.}
  \label{NH}
\end{figure}

The $\rm N_H$ values for all the sources in our sample are estimated using the calculated HRs. PIMMS tool \citep[Portable, Interactive, Multi-Mission Simulator;][]{Mukai1993} is used to create a grid of HR and $\rm N_H$ values and convert the HR values from BEHR into $\rm N_H$. PIMMs is also used to estimate the correction factor, that is applied for the calculation of the intrinsic fluxes of our sources (see next paragraph). In our calculations we assumed that the power law X-ray spectrum has a fixed photon index ($\Gamma$) of 1.8. The value of galactic $\rm N_H$ used is $\rm log  ~\rm N_{H,gal}/ ~cm^{-2}= 20.25$. The AGN sample is split into X-ray absorbed and unabsorbed sources, using a $\rm N_H$ cut at  $\rm log  ~\rm N_H/ ~cm^{-2}= 21.5$. This value has been chosen in a number of previous studies since it provides a good agreement between the X-ray and optical classification of type 1 and 2 AGNs  \citep[e.g.][]{Merloni2014, Masoura2020}. Fig. \ref{NH} presents the distribution of the hydrogen column density for the examined sources. Its bimodal nature, was observed in previous works, too \citep[e.g.][]{Civano2016, Stemo2020}.

\begin{figure*}
\centering
\begin{subfigure}{.505\textwidth}
  \centering
  \includegraphics[width=.92\linewidth]{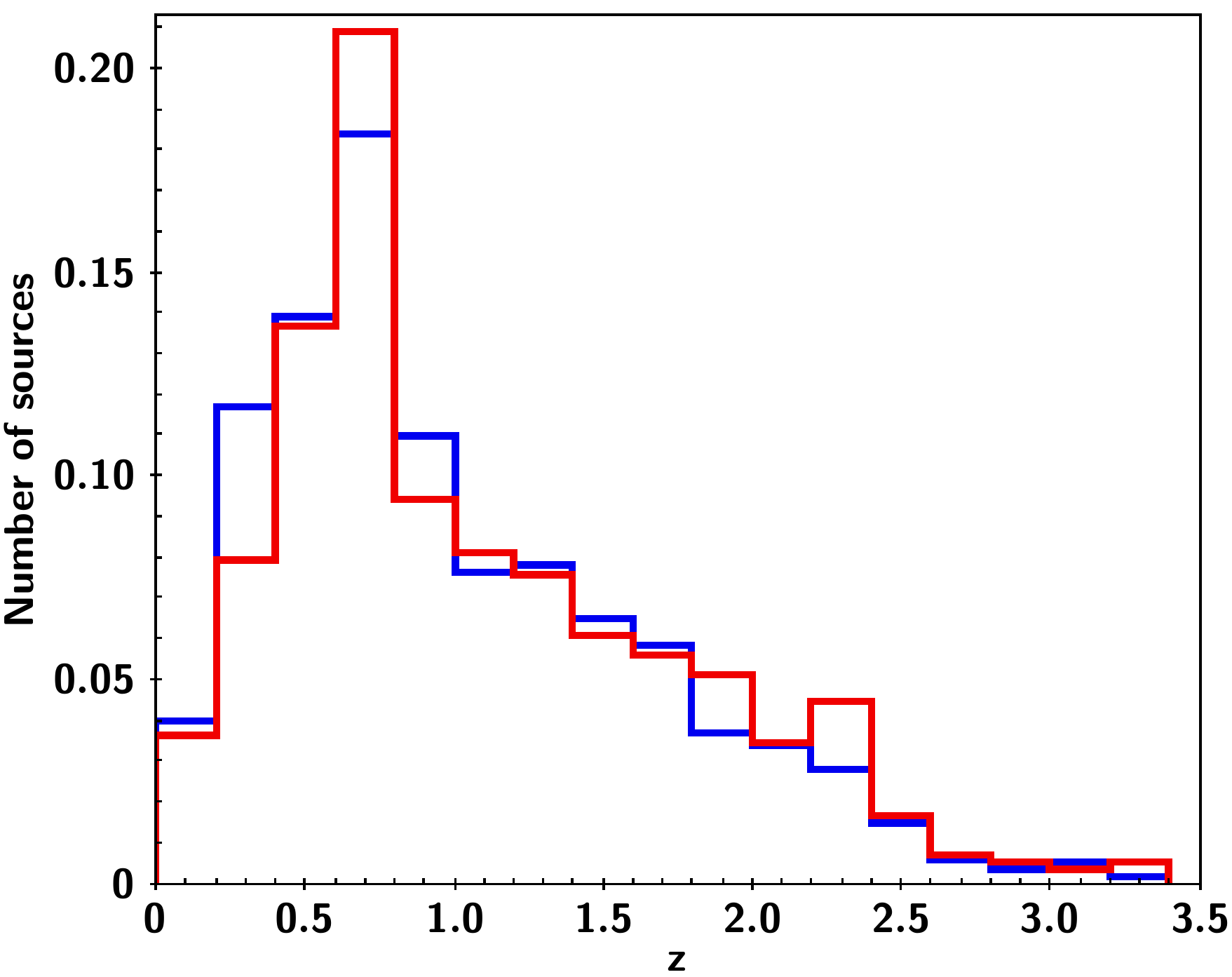}
  \label{z}
\end{subfigure}%
\begin{subfigure}{.52\textwidth}
  \centering
  \includegraphics[width=.92\linewidth]{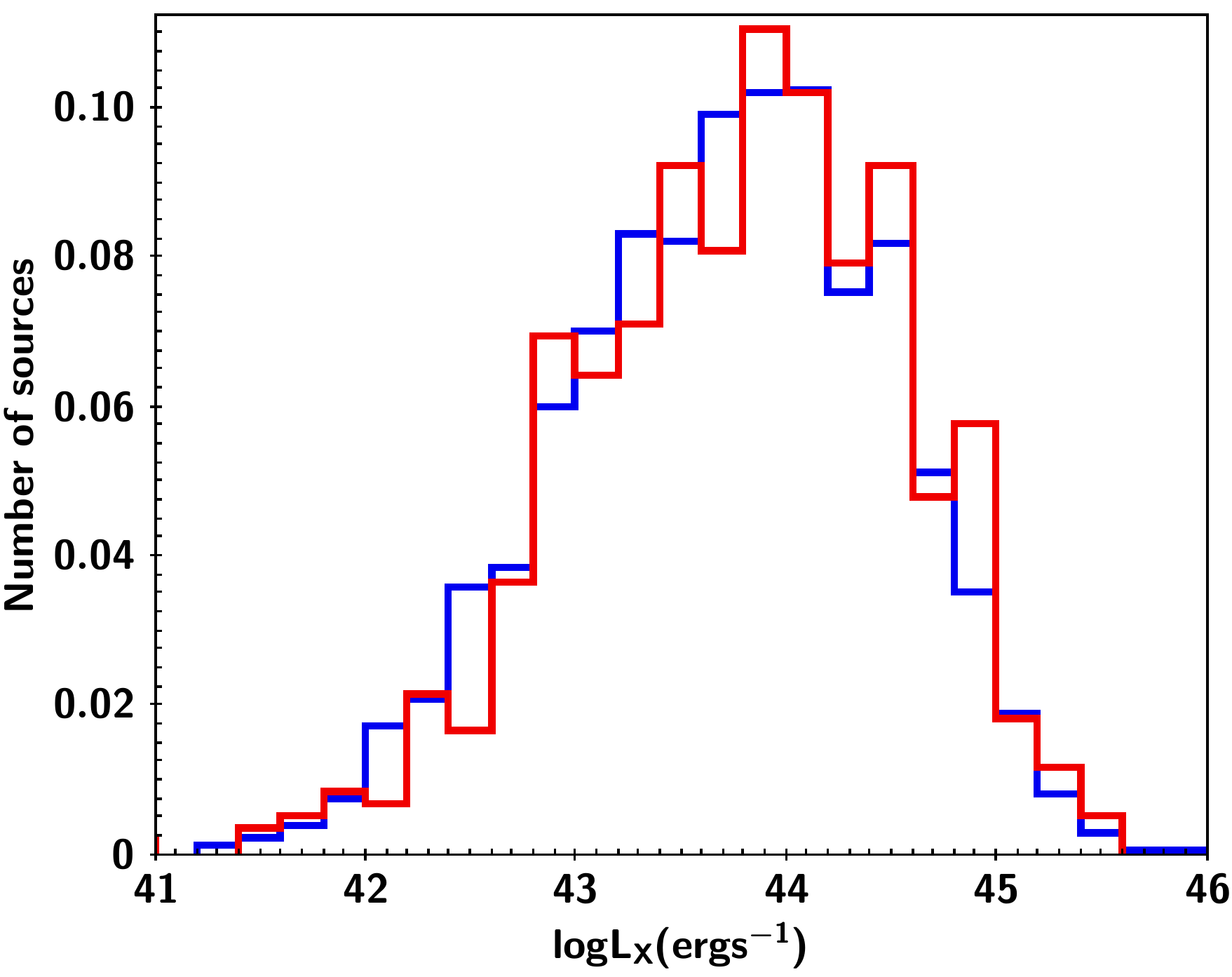}
  \label{Lx}
\end{subfigure}
\caption{Left: Redshift distribution of the examined sample. Right: Intrinsic, X-ray luminosity distribution of the examined sample. Blue and red colours refer to the unabsorbed and absorbed sources, respectively. Both histograms have been normalised to the total number of sources.}
\label{z_Lx}
\end{figure*}

Having estimated $\rm N_H$, we compute the correction factor, defined as $f_{int}/f_{abs}$, where $f_{abs}$ is the absorbed flux and $f_{int}$ the intrinsic flux. The latter has been estimated using PIMMS and assuming an unabsorbed power law with $\rm \Gamma =1.8$. Then, each observed X-ray luminosity is corrected using this factor, to estimate the intrinsic X-ray luminosity. Fig. \ref{z_Lx} presents the distribution of redshift (left panel) and intrinsic, X-ray luminosity (right panel) of absorbed (red line) and unabsorbed (blue line) AGN. We notice that the distributions of the two AGN subsamples are similar. This can be explained by the low X-ray absorption limit we adopt in our analysis. Adoption of a higher $\rm N_H$ cut ($\rm log  ~\rm N_H/ ~cm^{-2}= 22$), results in different redshift and X-ray luminosity distributions between absorbed and unabsorbed sources, in agreement with the predictions of X-ray luminosity function studies \citep[e.g.][]{Aird2015}.

\section{Results \& Discussion}
In this section, we present our main results and compare them with previous studies. Specifically, we examine, whether absorbed and unabsorbed AGN have different host galaxy properties and if the SFR-$\rm L_X$ relation changes with absorption.

\subsection{Host galaxy properties of absorbed and unabsorbed AGN}
\label{sec_weight}
As presented in Fig. \ref{z_Lx}, the redshift and $\rm L_X$ distributions of X-ray absorbed and unabsorbed classified AGN in our sample are similar. However, we account for the small differences between them, to facilitate a better comparison of the host galaxy properties of the two populations. For that purpose, we join the redshift distributions of the two populations and normalise each one by the total number of sources in each redshift bin (bin size of 0.1). We repeat the same process for the $\rm L_X$ distributions of the two subsamples (bin size of 0.2\,dex). This procedure provides us with the probability density function (PDF) in this 2$-$D parameter space (i.e., redshift and $\rm L_X$ space). Then, we use the redshift and $\rm L_X$ of each source to weigh it based on the estimated PDF \citep[see also][]{Mendez2016, Mountrichas2016}. This correction method is similar to that applied in previous studies \citep[e.g.][]{Zou2019} and allows a fair comparison with their results. Additionally, each source is weighted based on the statistical significance of its stellar mass and SFR measurements (see next Sections for details).

\subsubsection{Stellar mass distribution}
\label{mstar}

\begin{figure}
\centering
  \includegraphics[width=1.\linewidth, height=7.cm]{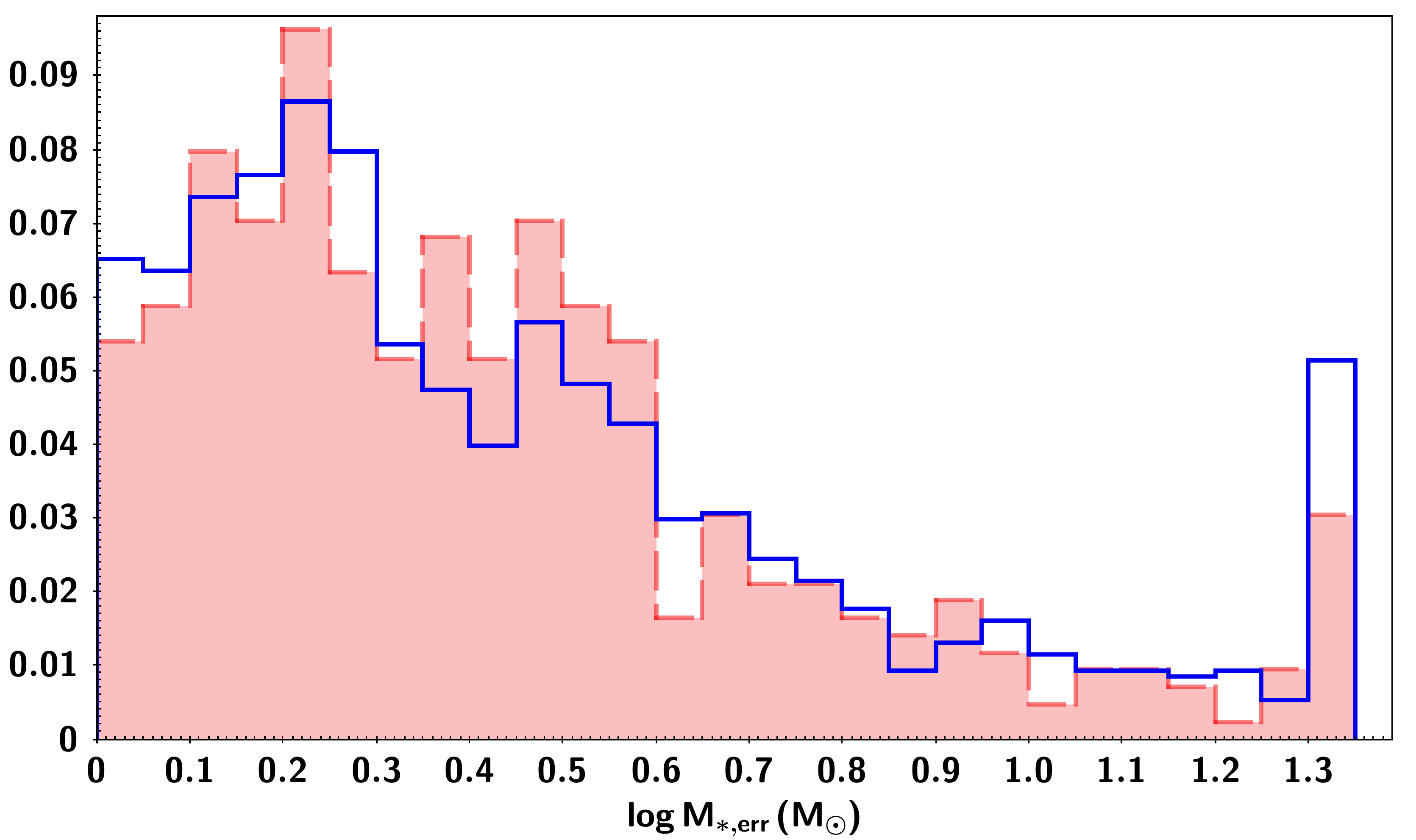}
      \caption{Distributions of the error of the log\, $\rm M_*$ for absorbed (red shaded histogram) and
unabsorbed (blue line) AGN. The two populations have similar error distributions, with the
bulk of the measurements to have errors $\rm \leq0.5-0.6\,dex$.}
\label{stellar_mass_err}
\end{figure}

Stellar mass estimations for AGN may suffer from large uncertainties, in particular in the case of unabsorbed sources. The AGN emission can outshine the optical emission of the host galaxy, thus rendering stellar mass calculations uncertain. The median stellar mass measurements and the median uncertainties, estimated by CIGALE are $10.6\pm0.4$ and  $10.9\pm0.6$, for unabsorbed and absorbed AGN, respectively. Restricting our sample to the most X-ray luminous ($\rm L_X \geq 10^{44}$ $\rm erg s^{-1}$), unabsorbed AGN ($\sim 20\%$ of our total sample) gives $11.1\pm0.8$. Fig. \ref{stellar_mass_err} presents the distributions of stellar mass uncertainties for absorbed (red shaded area) and unabsorbed (blue line) AGN. The two populations have similar distributions, with the bulk of the measurements to have errors $\rm \leq0.5-0.6\,dex$. We choose not to exclude sources with large stellar mass errors to avoid reducing the size of the dataset. Instead we account for the M$_*$ uncertainties, by estimating the significance (value/uncertainty), $\sigma$, of each stellar mass measurement and weigh each source based on its $\sigma$ value (in addition to the weight presented in the previous paragraph). We confirm, that excluding from our analysis those sources that have stellar mass uncertainties larger than 0.6\,dex ($\sim 25\%$ of unabsorbed and $\sim 33\%$ of absorbed AGN) does not change our results and conclusions compared to those presented, using the weighting method described above.

\begin{figure}
\centering
  \includegraphics[width=0.94\linewidth]{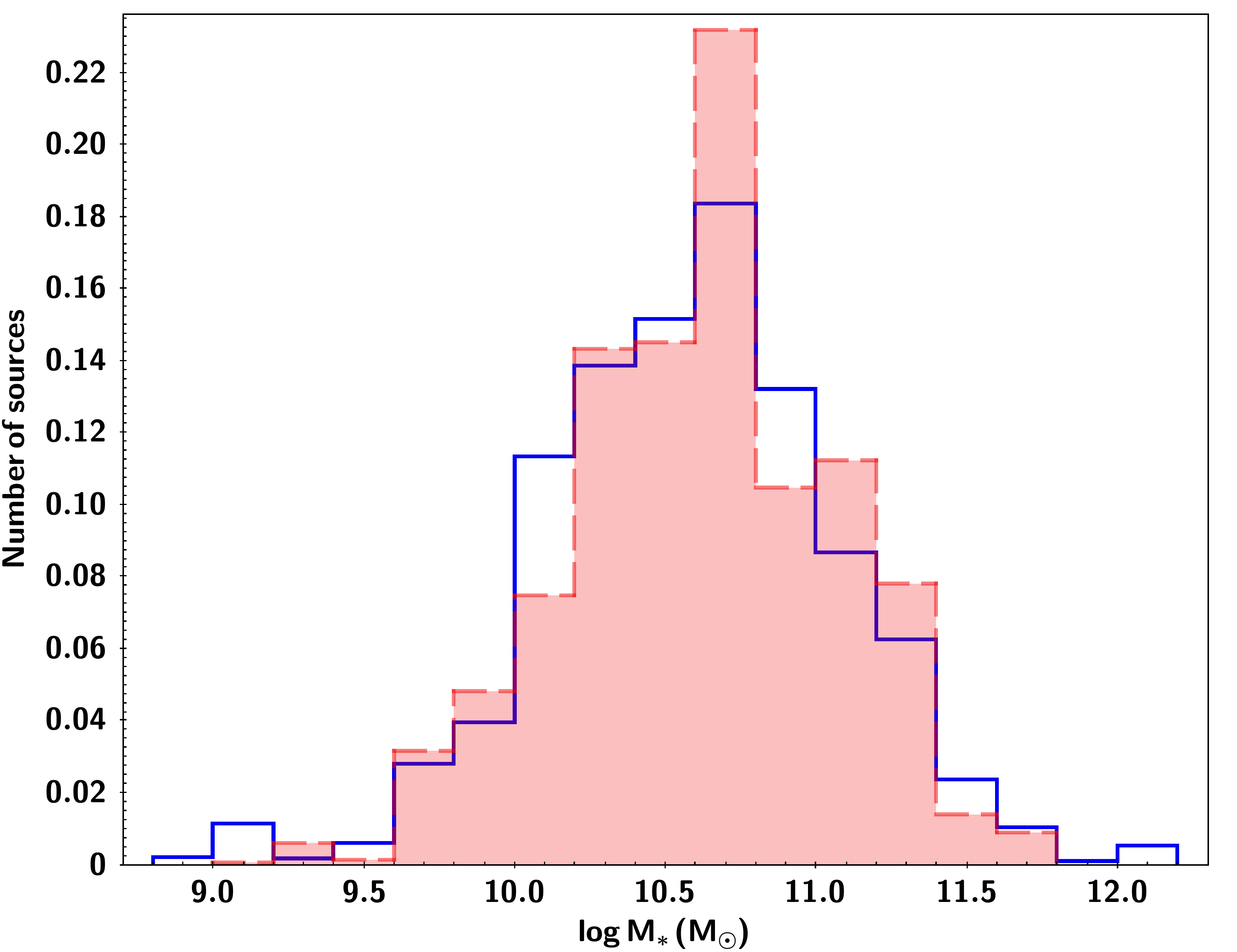}
      \caption{Stellar mass distributions. Blue and red colours refer to the unabsorbed and absorbed sources, respectively. Histogram has been normalised to the total number of sources. Based on the KS-test ($p-\rm value=0.72$), the two AGN populations have similar M$_*$ distributions.}
\label{stellar_mass}
\end{figure}

Fig. \ref{stellar_mass} presents the distribution of M$_*$ for both absorbed and unabsorbed AGNs, in bins of 0.2\,dex. We apply a two-sample Kolmogorov-Smirnov test (KS-test) to examine whether or not the two distributions differ. The KS-test reveals that the distributions are similar for both AGN populations ($p-\rm value=0.72$). We also split the AGN sample into low and high redshift subsamples, using a redshift cut at $\rm z=1.0$ and repeat the process. KS-test shows no statistical significant difference of the M$_*$ distribution of absorbed and unabsorbed sources, at any redshift. Specifically, at $z<1.0$ KS-test gives $p=0.82$ and at $z>1.0$, $p=0.39$. Our findings are in agreement with \cite{Merloni2014} who used 1310 X-ray selected AGN from the XMM-COSMOS survey with redshift range 0.3 < z < 3.5. In that study, the X-ray population was split into obscured and unobscured AGN, using the same N$_H$ cut that is applied in our analysis (N$_H$=$10^{21.5}\, \rm{cm^{-2}}$). They found no remarkable differences between the mean M$_*$ of obscured and unobscured AGNs. On the other hand, \cite{Lanzuisi2017}, used approximately 700 X-ray selected AGNs, in the COSMOS field in the redshift range 0.1 < z < 4, and found that unobscured AGNs tend to have lower M$_*$ than obscured sources. However, they considered as X-ray absorbed sources with $\rm N_H > 10^{22}\, \rm{cm^{-2}}$. To be consistent with \cite{Lanzuisi2017}, we apply the KS-test adopting the same N$_H$ with them. The estimated $p$-value, although is reduced to 0.56 still indicates that the M$_*$ distribution is similar, for the two AGN populations. Furthermore, \cite{Zou2019} used 2463 X-ray selected AGNs in the COSMOS field and found that unobscured AGNs tend to have lower M$_*$ than their obscured counterparts. However, in their analysis they divided their sample into type 1 and type 2 AGNs based on their optical spectra, morphologies, and variability. Thus, the disagreement with our results could be attributed to the different method applied in the characterisation of a source as obscured/unobscured.  

\begin{figure*}
\centering
\begin{subfigure}{.505\textwidth}
  \centering
  \includegraphics[width=0.99\linewidth]{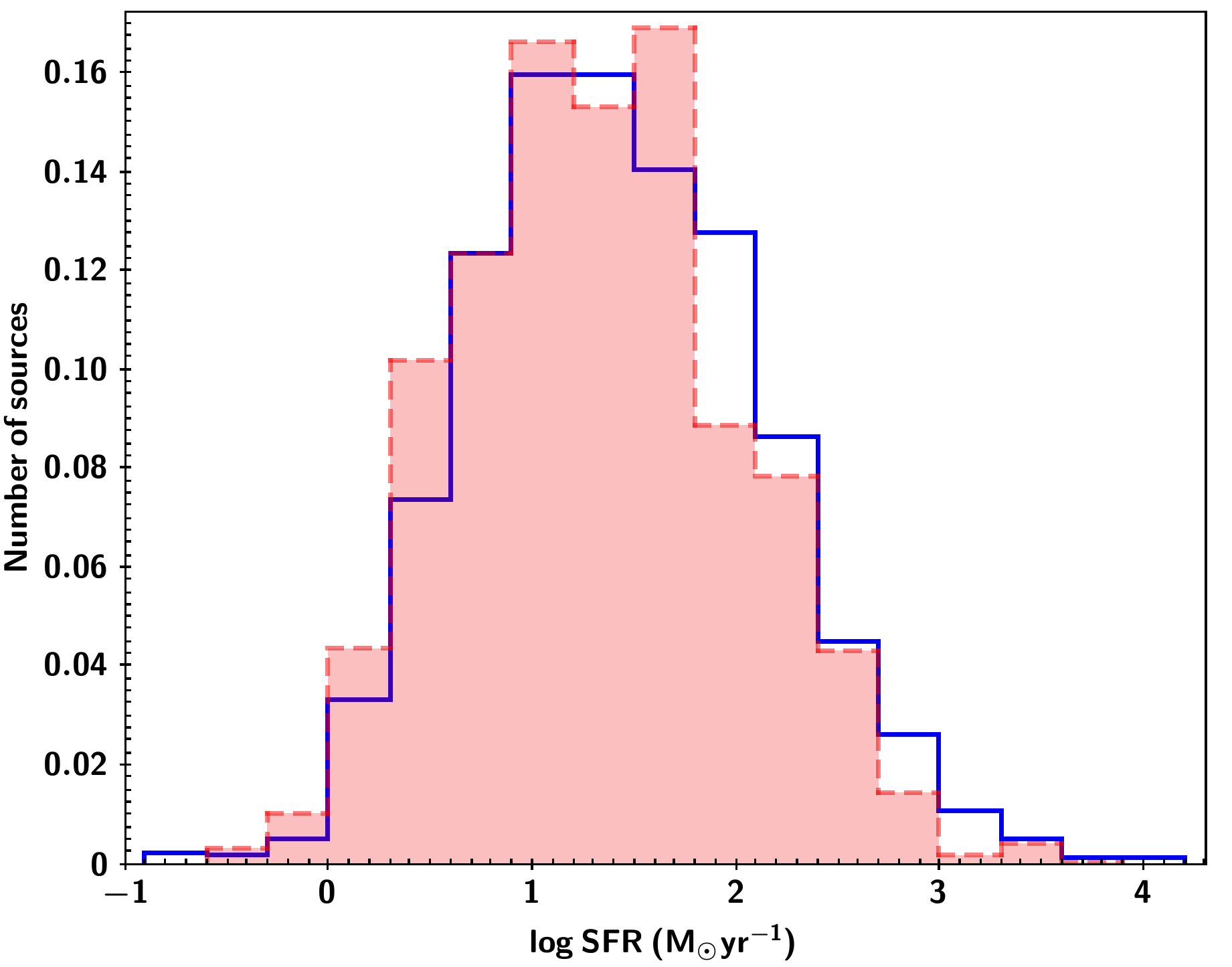}
\end{subfigure}%
\begin{subfigure}{.52\textwidth}
  \centering
  \includegraphics[width=.92\linewidth]{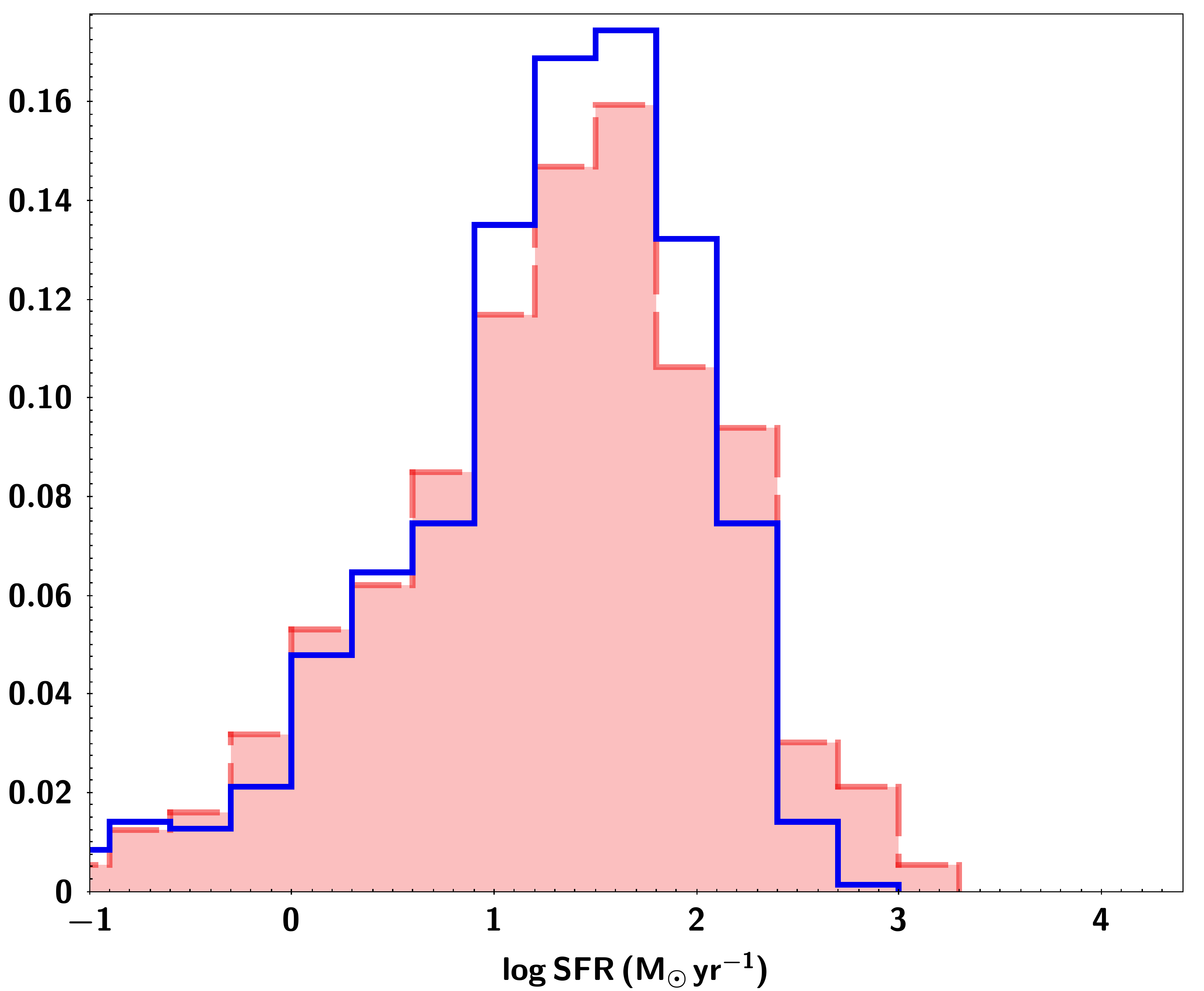}
\end{subfigure}
      \caption{Star formation rate distribution. Blue and red colours refer to the unabsorbed and absorbed sources, respectively. Histograms have been normalised to the total number of sources. Left: Distributions for the total number of sources. The two populations have similar SFR distributions ($p-\rm value=0.95$). Right: Distributions for the 1,276 AGN with Herschel photometry available. We confirm that the X-ray absorbed and unabsorbed AGN have similar SFR distributions ($p-\rm value=0.86$).}
\label{SFR}
\end{figure*}

\begin{figure}
\centering
  \includegraphics[width=0.958\linewidth]{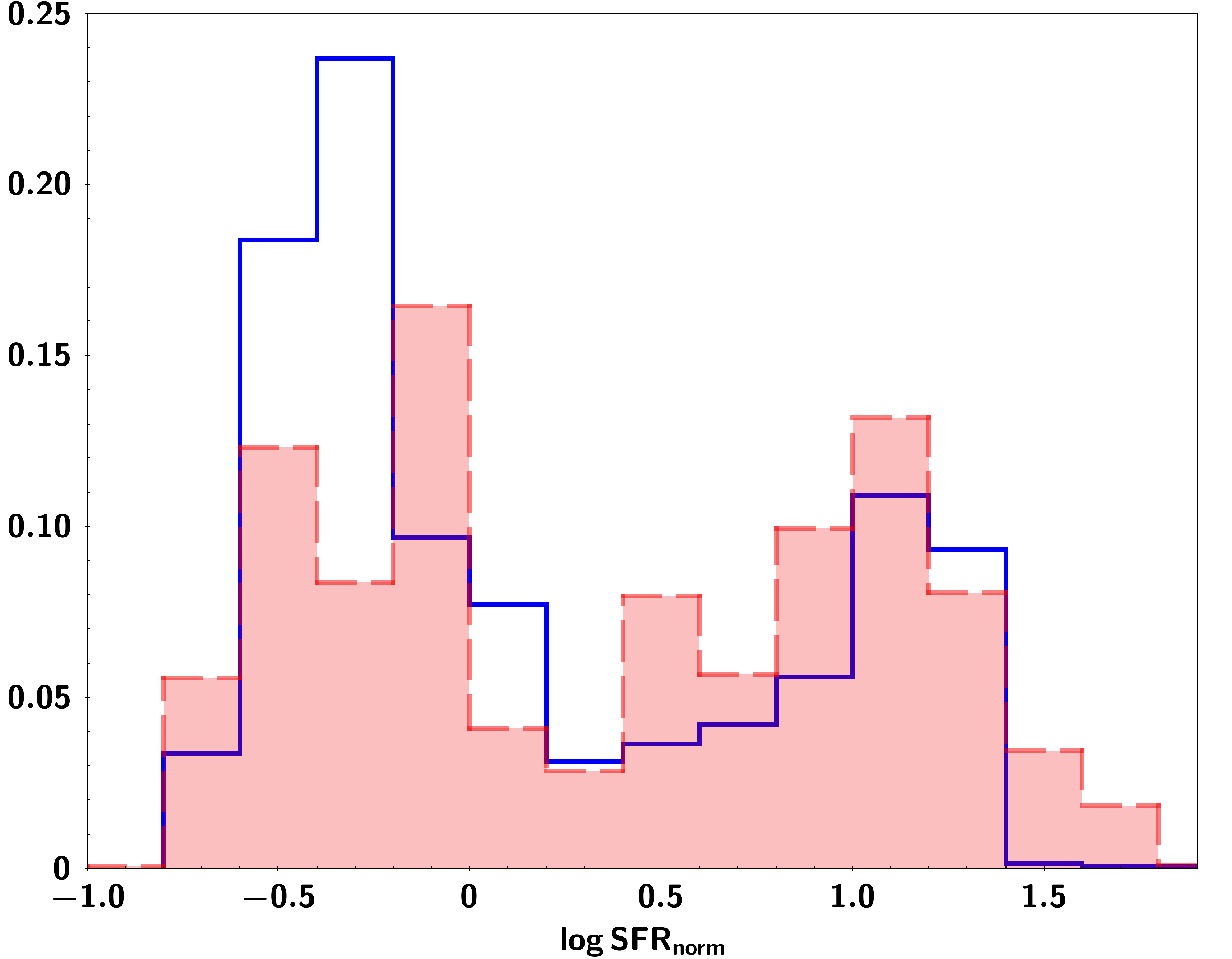}
      \caption{Normalized star formation rate distribution. Blue and red colours refer to the unabsorbed and absorbed sources, respectively. The histogram has been normalised to the total number of sources. The two populations have similar normalised SFR distributions ($p-\rm value=0.71$). Both distributions are bimodal.}
\label{SFR_norm}
\end{figure}

\begin{figure}
\centering
  \includegraphics[width=0.958\linewidth]{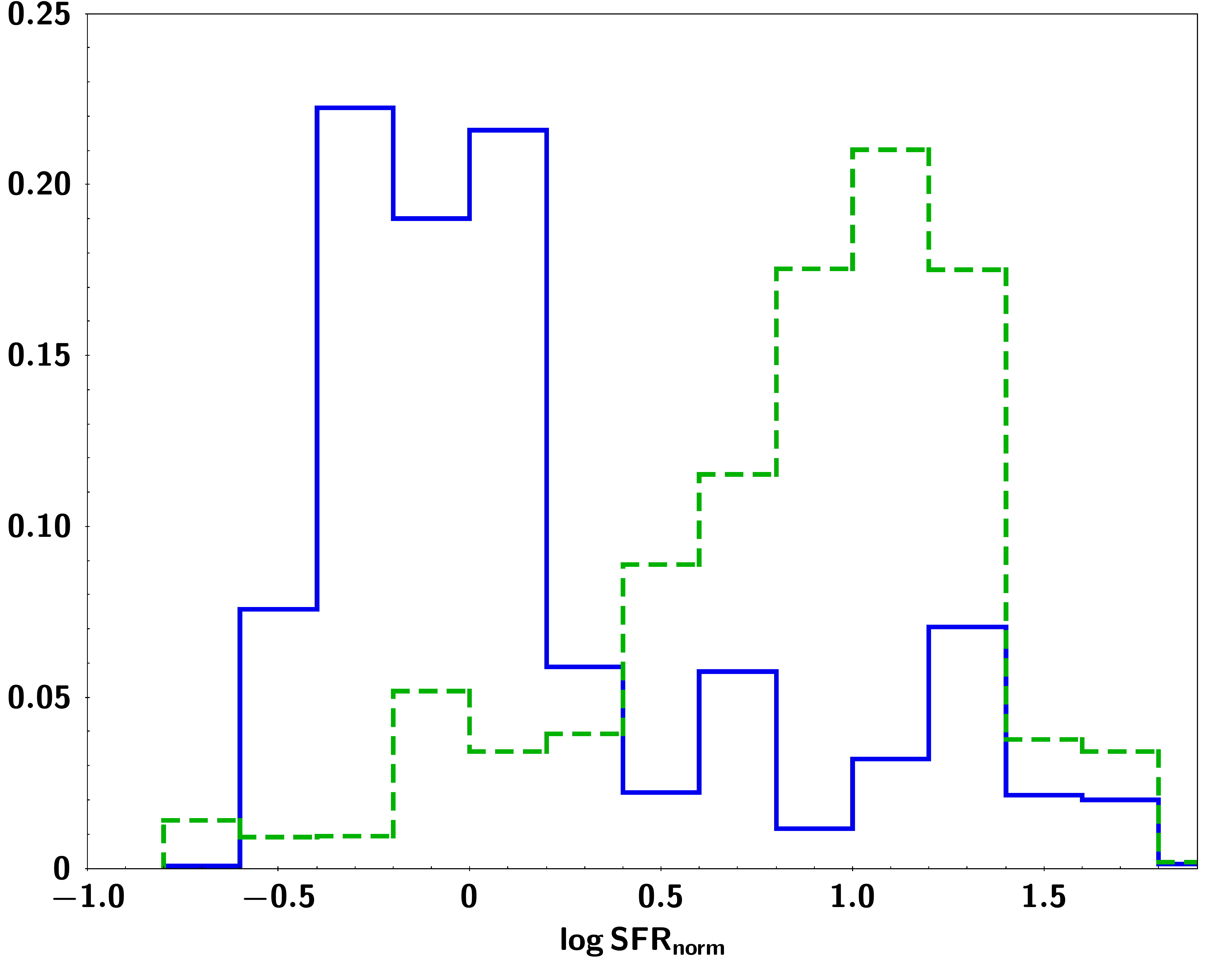}
      \caption{Normalized star formation rate distribution for sources at $\rm z<1.2$ (blue line) and at $\rm z>1.2$ (green, dashed line). The histogram has been normalised to the total number of sources. AGN in the lower redshift bin have SFR comparable with galaxies in the star-forming main sequence ($\rm log\,SFR_{{\it norm}}=0$), whereas AGN at higher redshift have enhanced SFR compared to galaxies in the main sequence. This suggests evolution of the SFR$_{norm}$ with redshift (see text for more details).}
\label{SFR_norm_zbins}
\end{figure}

\begin{figure}
\centering
  \includegraphics[width=0.95\linewidth]{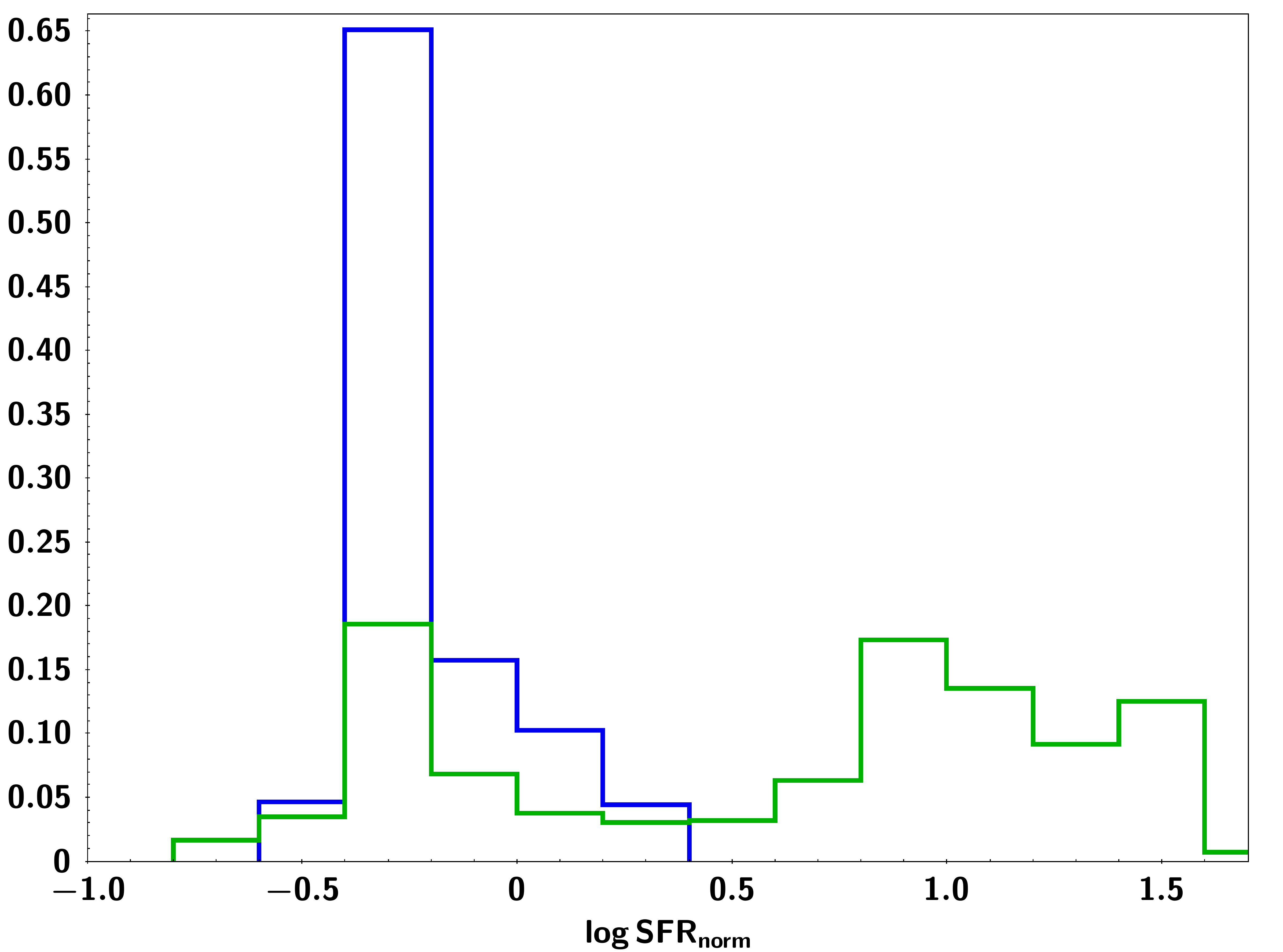}
\caption{SFR$_{norm}$ distribution for low ($\rm L_X < 2 \times 10^{43}$ $\rm erg s^{-1}$, blue line) and high ($\rm L_X > 2 \times 10^{43}$ $\rm erg s^{-1}$, green line) X-ray luminosities. Redshift and $\rm L_X$ range have been restricted to match that of \cite{Bernhard2018}. We only present measurements for the X-ray absorbed AGN.}
\label{comparison_bern}
\end{figure}

\subsubsection{Star formation rate distribution}
\label{star}

The left panel of Fig. \ref{SFR} presents the distributions of the SFR, for both absorbed and unabsorbed AGNs, in bins of 0.3\,dex for the total sample. Each source is weighted based on its X-ray luminosity and redshift as well as the $\sigma$ value of its SFR measurement. Application of the two-sample KS-test reveals that the SFR distributions are similar for both AGN populations, with $p-\rm value$ equals to 0.95. This is also the case when we split the AGN sample into low and high redshift subsamples, using a redshift cut at $\rm z=1.0$ (for $z<1.0$, $p=0.99$ and for  $z>1.0$, $p=0.86$). In the right panel, we present the SFR distributions only for those AGN that have available {\it Herschel} photometry. The results are similar to those for the full sample, i.e. the SFR distributions of X-ray absorbed and unabsorbed AGN present no significant differences ($p-\rm value=0.86$). Further restricting the sample to those sources with signal-to-noise ratio greater than 3 for the SPIRE bands (216 AGN), does not change the results.

Our findings agree with most previous studies \citep{Merloni2014, Lanzuisi2017, Zou2019}. On the other hand, \cite{Chen2015} used AGN from the B\"ootes field and claimed that type 2 sources have higher IR star formation luminosities (a proxy of star formation) by a factor of $\sim2$ than type 1. However, their dataset consists of mid-IR selected, luminous quasars. Moreover, they have divided their sample into type 1 and type 2 AGNs using optical/mid-IR colour criteria R$-$[4.5] = 6.1 \citep[e.g.][]{Hickox2007}, where R and [4.5] are the Vega magnitudes in the R and IRAC 4.5$\mu m$ bands, respectively.

\subsubsection{Normalised star formation rate distribution}
\label{sec_sfr_norm}

We calculate the SFR$_{norm}$, to take into account the evolution of SFR with stellar mass and redshift (see section \ref{prop}). The distribution of SFR$_{norm}$ of absorbed and unabsorbed AGN is presented in Fig. \ref{SFR_norm}, in bins of 0.2\,dex. Each source is weighted based on its redshift and luminosity, as described in Section \ref{sec_weight}, as well as based on the significance of its stellar mass and star formation rate measurement. The KS-test results in a $p$-value equal to 0.71, i.e., the two populations have similar SFR$_{norm}$ distributions. We notice that both distributions are bimodal. Further investigation reveals that, both for the absorbed and the unabsorbed sources, the first peak is due to low redshift systems (mean $\rm z\approx0.7$), while the second peak is due to AGN that lie at high redshifts (mean $\rm z\approx1.5$). This may suggest evolution of the SFR$_{norm}$ with redshift. To investigate this scenario further, we split the AGN sample into two redshift bins, at $\rm z=1.2$. The SFR$_{norm}$ distributions are presented in Fig. \ref{SFR_norm_zbins}. The results confirm that the SFR$_{norm}$ distribution peaks at different values at low and high redshifts. AGN in the lower redshift bin have SFR comparable with galaxies in the star-forming main sequence ($\rm log\,SFR_{{\it norm}}=0$), whereas AGN at higher redshift have enhanced SFR compared to galaxies in the main sequence. We discuss this in more detail in the next section.

\cite{Bernhard2018} examined the SFR distribution of X-ray AGN, taking into consideration its evolution with M$_*$ and redshift. Specifically, they used X-ray sources in the COSMOS field and investigated the SFR$_{norm}$ distribution as a function of $\rm L_X$. Based on their analysis, more powerful AGN present a narrower SFR$_{norm}$ distribution that peaks close to that of the MS galaxies. Their sample consists of AGN at intermediate redshifts, i.e., $\rm 0.8<z<1.2$. Based on the mean redshift values of the systems contained in the two peaks of our SFR$_{norm}$ distributions presented above, this would place their sources in between our two peaks and close to SFR of MS galaxies. Furthermore, Bernhard et al. sample spans a narrower $\rm L_X$ range and lacks sources both at low and in particular at high $\rm L_X$ compared to the sample used in our analysis (see our Fig. \ref{Lx_z_1} and their Fig. 1). There are only $\sim 500$ AGN in our dataset that are within the redshift and luminosity range of Bernhard et al. Further restricting the sample to only obscured sources, in accordance with Bernhard et al., results in less than 200 sources. Applying their luminosity cut ($\rm L_X = 2 \times 10^{43}$ $\rm erg s^{-1}$) we divide AGN into high and low luminosity systems. We note that there are less than 30 low luminosity sources in our sample. The SFR$_{norm}$ distribution of high (green) and low (blue) $\rm L_X$ AGN is presented in Fig. \ref{comparison_bern}. High luminosity AGN still present a two-peaked SFR$_{norm}$. We note, however, that although we have matched the redshift and $\rm L_X$ range with that of Bernhard et al., the $\rm L_X$ distributions of the two datasets are still different. Nevertheless, the two results agree that higher X-ray luminosity AGN exhibit higher SFR$_{norm}$ values. We conclude, that the different results between our study and Bernhard et al. could be due to the different  $\rm L_X$, redshift plane probed by the samples of the two studies and in particular the different luminosity distributions between the two datasets.

\subsection{$\rm SFR_{norm}-\rm L_X$ correlation for different AGN types}
\label{sec_sfrnorm_lx}

In this section, we study the effect of obscuration on the SFR$-\rm L_X$ relation. To account for the evolution of SFR with stellar mass and redshift (see also section \ref{prop}), we plot SFR$_{norm}$ estimations versus $\rm L_X$. 

Luminous sources are observed within a larger volume compared to their faint counterparts, in flux limited surveys (Fig. \ref{Lx_z_1}). This introduces a selection bias to our analysis. To account for this effect, we adopt the $\rm \frac {1}{V_{max}}$ correction method \citep{Akylas2006, Page2000}. Specifically, we estimate for each source of a given $\rm L_X$, the maximum available volume that it can be observed, using the following equation:
\begin{equation}
\rm V_{max} = \int_{0}^{\rm z_{max}} \Omega(f) \rm \frac {dV}{dz} dz,
\end{equation}
where $\Omega(f)$ is the value of the sensitivity curve at a given flux, corresponding to a source at a redshift z with observed luminosity $\rm L_X$. $\rm z_{max}$ is the maximum redshift at which the source can be observed at the flux limit of the survey. The area curve used in our calculations is presented in \cite{Liu2016} (see their Fig. 3). Therefore, each source is weighted by the $\rm \frac {1}{V_{max}}$ value, depending on their $\rm L_X$ and redshift. An additional weight is also considered, based on the $\sigma$ value of the M$_*$ and SFR measurements.

\begin{figure*}
\centering
\begin{subfigure}{.505\textwidth}
  \centering
  \includegraphics[width=.9\linewidth]{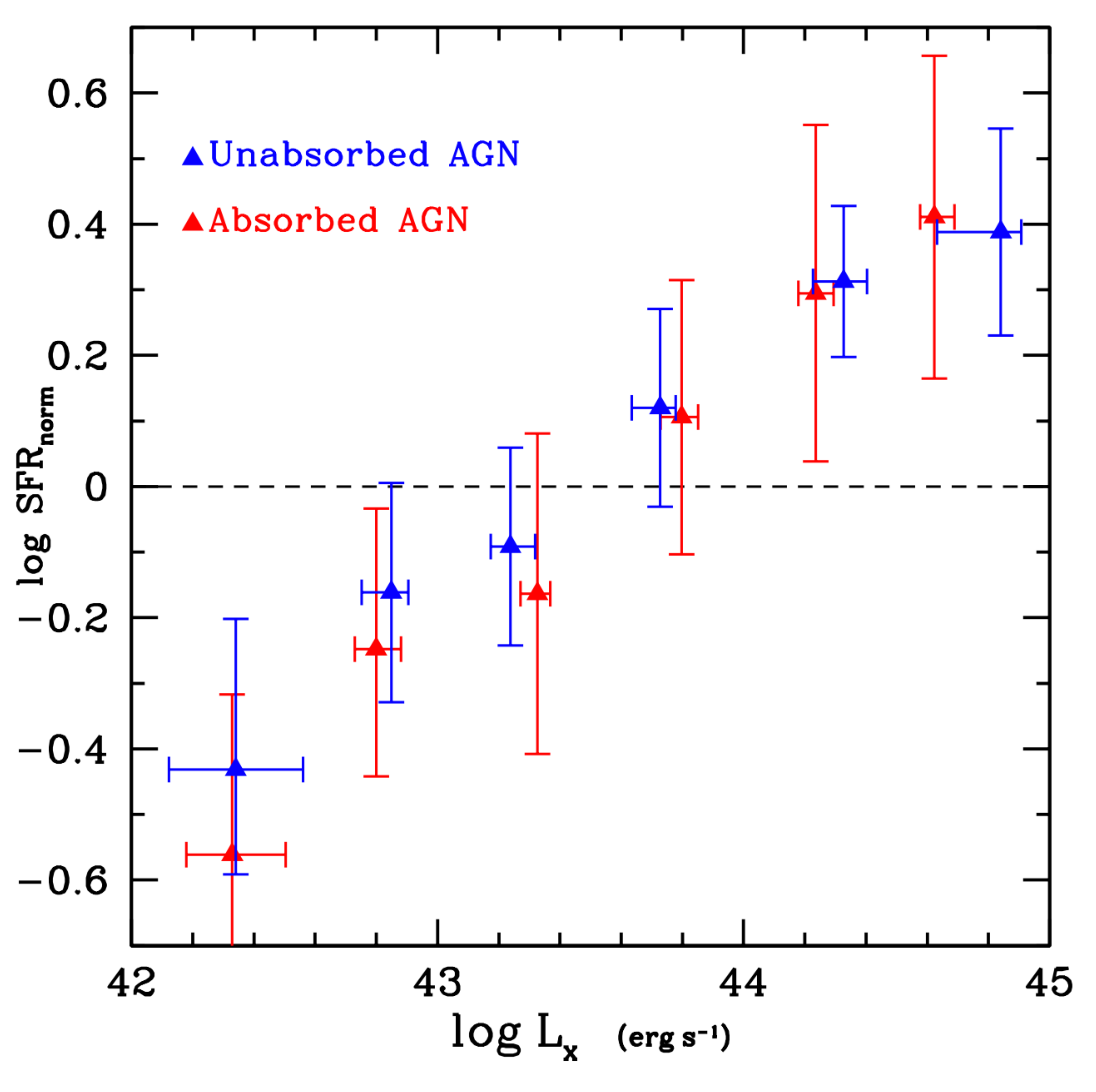}
 \label{lxbinabs}
\end{subfigure}%
\begin{subfigure}{.505\textwidth}
  \centering
  \includegraphics[width=.9\linewidth]{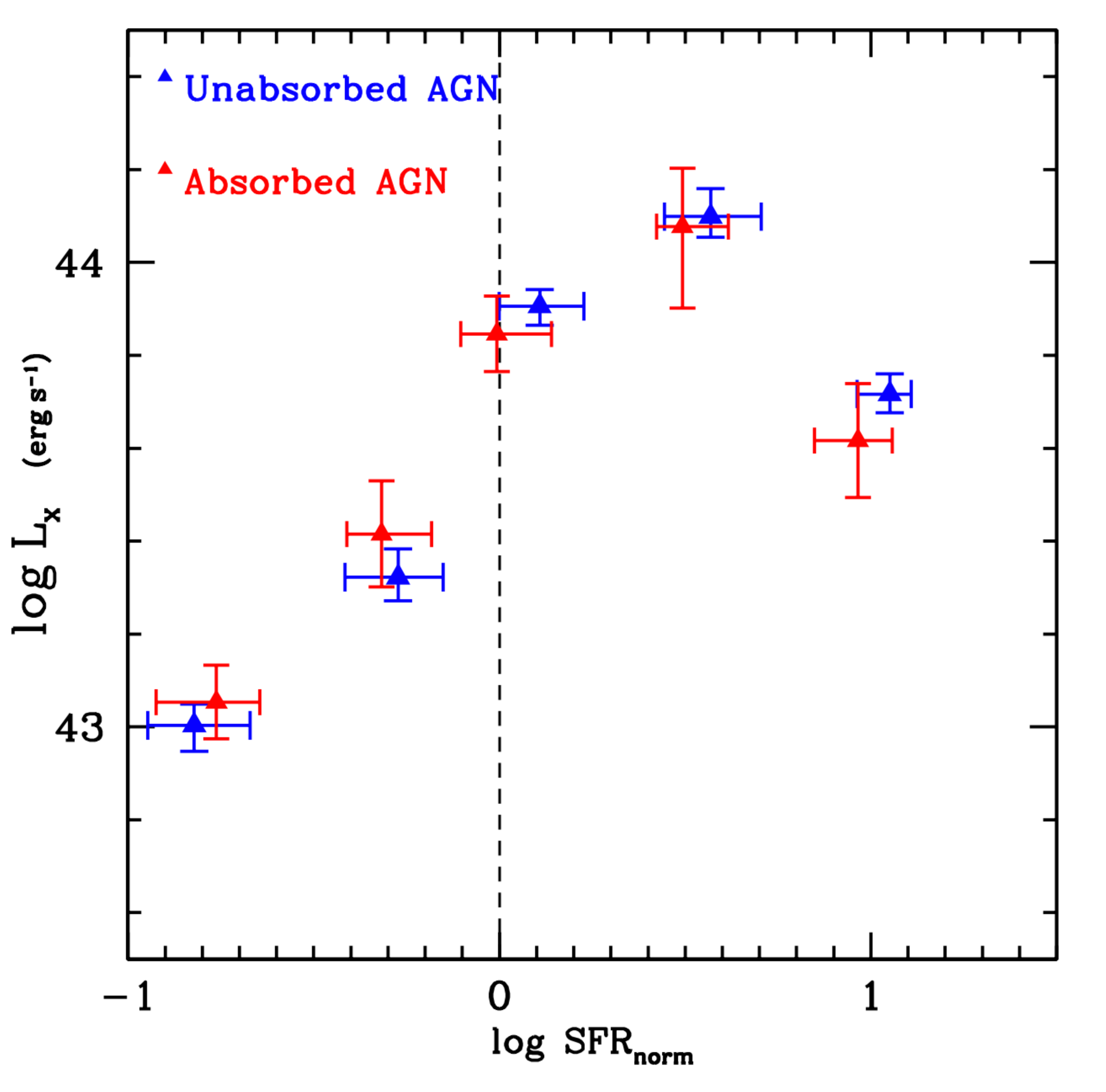}
  \label{sfrbinabs}
\end{subfigure}
\caption{Left: The SFR$_{norm}$$-$$\rm L_X$ correlation in $\rm L_X$ bins, for absorbed and unabsorbed sources. Based on our measurements, the AGN enhances the SFR of the host galaxy, at all X-ray luminosities spanned by our sample, regardless of whether the source is X-ray absorbed or not. Right: The SFR$_{norm}$$-$$\rm L_X$ correlation in SFR$_{norm}$ bins, for absorbed and unabsorbed sources \citep{Masoura2018}. Absorbed and unabsorbed sources follow, approximately, the same trend. In both plots, red and blue triangles refer to the absorbed and unabsorbed sources, respectively. The dashed line corresponds to the star forming main sequence.  Median SFR and $\rm L_X$ values are presented, in both panels. The error bars represent the 1$\sigma$ dispersion of each bin. Based on our measurements, the SFR$_{norm}$$-$$\rm L_X$ correlation is similar for different AGN types.}
\label{sfr_lx}
\end{figure*}

In Fig. \ref{sfr_lx} (left panel), we present our measurements in $\rm L_X$ bins, for absorbed and unabsorbed AGN, in bins of size 0.5\,dex. Median SFR and $\rm L_X$ values are presented. The error bars represent the 1\,$\sigma$ dispersion of each bin, i.e., they do not take into account the errors of the individual SFR and stellar mass measurements. The number of sources varies in each bin, but our results are not sensitive to the choice of the number and size of bins. Based on our findings, both X-ray absorbed and unabsorbed AGN present a similar SFR-$\rm L_X$ relation, at all X-ray luminosities spanned by our sample. Complementary, we present our measurements in SFR$_{norm}$ bins (bin size 0.5\,dex, right panel of Fig. \ref{sfr_lx}). Previous studies have also found that different binning affects the observed trends \citep[e.g.,][]{Lanzuisi2017, Hickox2014}. However, further examination of the underlying reasons that could be responsible for this effect are beyond the scope of this paper. Based on our findings, X-ray absorption does not play a significant role in the SFR$-\rm L_X$ relation, regardless of whether our measurements are binned in $\rm L_X$ or SFR bins.

\begin{figure*}
\centering
\begin{subfigure}{.33\textwidth}
  \centering
  \includegraphics[width=1.035\linewidth]{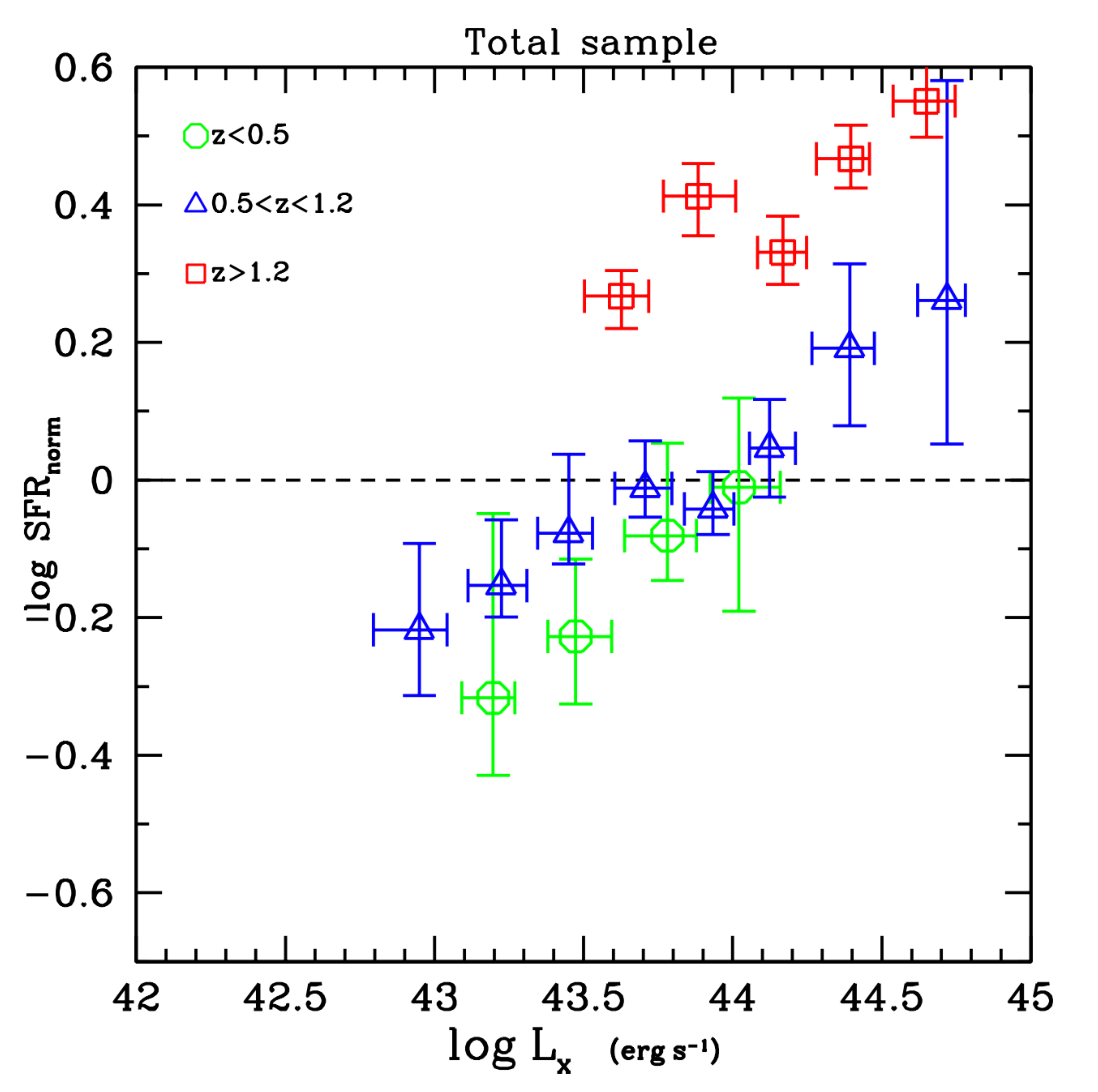}
 \label{lxbinabs}
\end{subfigure}
\begin{subfigure}{.33\textwidth}
  \includegraphics[width=1.035\linewidth]{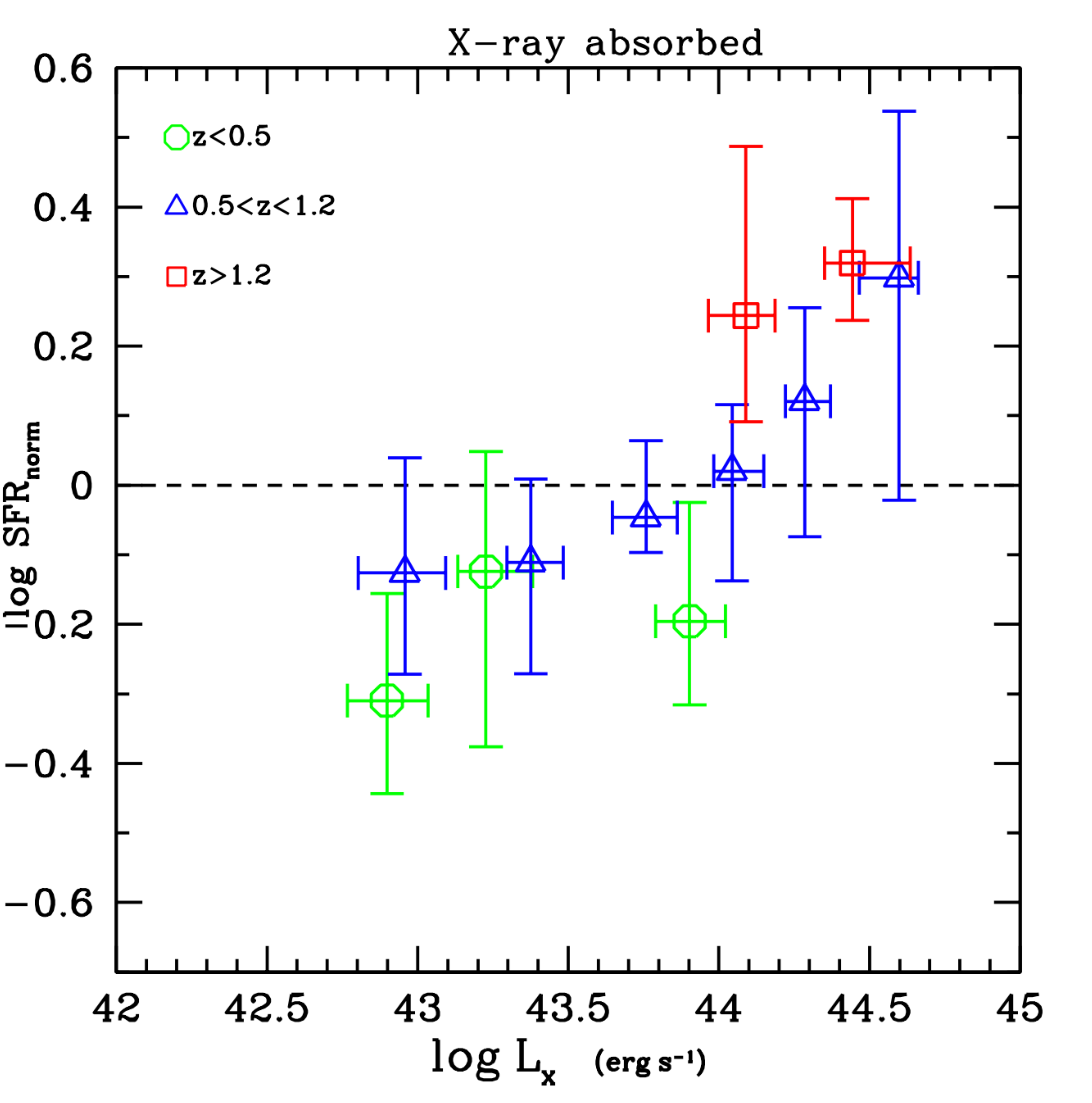}
 \label{lxbinabs}
\end{subfigure}
\begin{subfigure}{.33\textwidth}
  \includegraphics[width=1.035\linewidth]{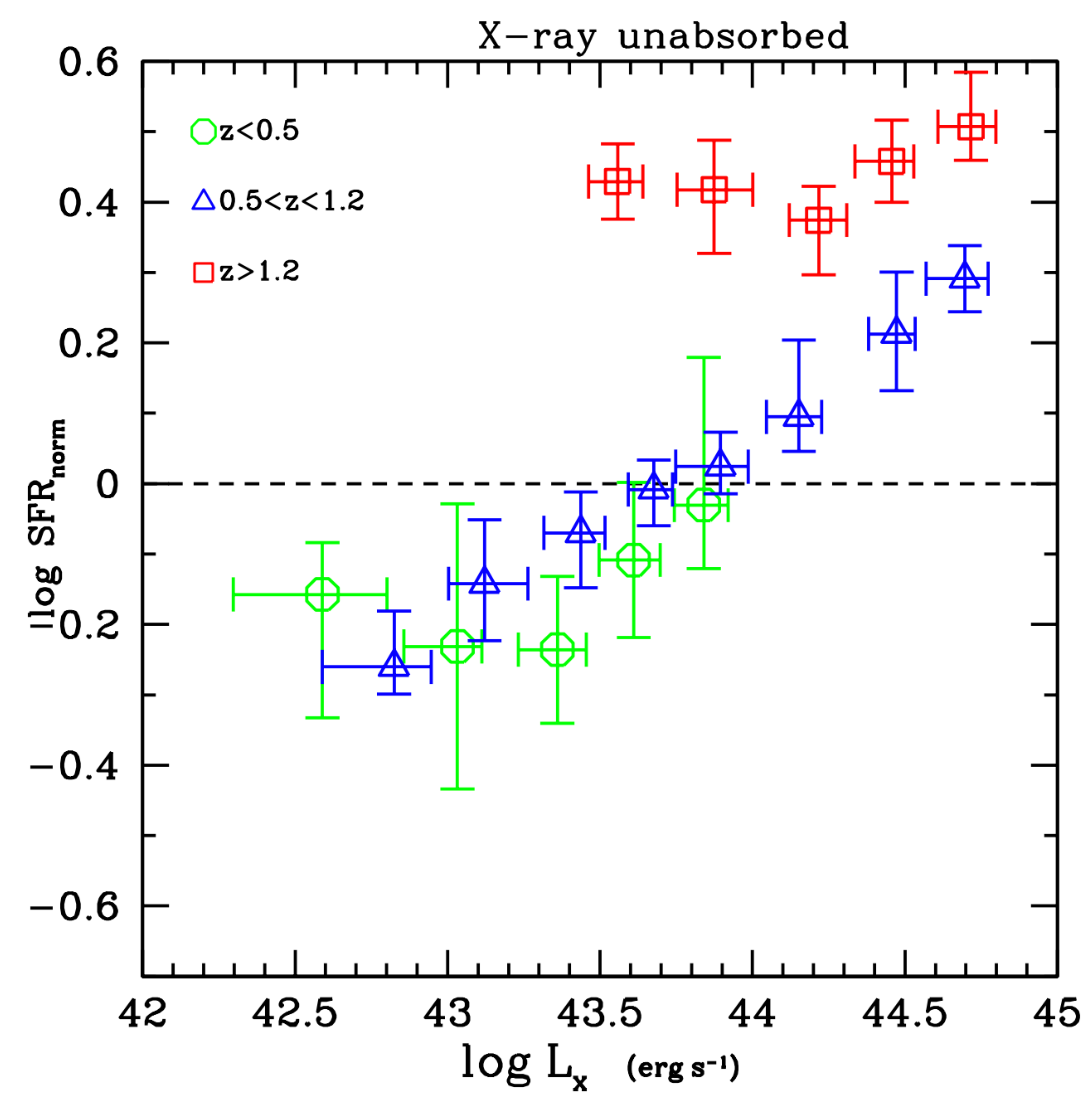}
 \label{lxbinabs}
\end{subfigure}
\caption{The SFR$_{norm}$$-$$\rm L_X$ correlation in $\rm L_X$ bins, for different redshift intervals ($\rm z<0.5$, $\rm 0.5<z<1.2$ and $\rm z>1.2$). Left, middle and right panels refer to the full sample, absorbed population and unabsorbed population, respectively. The dashed line corresponds to the star forming main sequence. Trends are similar in all redshift bins.  Median SFR and $\rm L_X$ values are presented. The error bars represent the 1\,$\sigma$ dispersion of each bin.}
\label{comparison_bins}
\end{figure*}

Additionally, we examine, if and how the redshift range affects our estimations. The $\rm N_H$ values, estimated in section \ref{abs_section} could be considered less secure, as we move to higher redshifts. This is because the absorption redshifts out of the soft-Xray band in the observed frame. Moreover, as shown in Fig. \ref{Lx_z_1} our sample lacks low-luminosity sources at high redshifts ($\rm z>1$). Thus, in Fig. \ref{comparison_bins}, we plot the SFR$_{norm}-\rm L_X$, for the whole sample (left), the absorbed (middle) and unabsorbed (right) subsamples using three redshift bins ($\rm z<0.5, 0.5<z<1.2\,and\,z>1.2$). Results are presented in $\rm L_X$ bins. The size of the bins varies from 0.25-0.5\,dex depending on the available number of sources of each subsample. Median SFR and $\rm L_X$ values are shown. The error bars represent the 1\,$\sigma$ dispersion of each bin. For X-ray absorbed sources we find no difference in the dependence of the SFR on the AGN power, at all redshifts. However, X-ray unabsorbed sources at high redshift ($\rm z>1.2$) present a flat SFR$_{norm}-\rm L_X$ relation.

Our results also show evolution of the SFR$_{norm}$ with redshift, for the full sample, as well as for X-ray absorbed and unabsorbed AGN (also seen in Fig. \ref{SFR_norm}). This evolution does not appear statistical significant for redshift below $\rm z<1.2$ ($\leq 1\,\sigma$), but its statistical significance increases ($\approx 2-2.5\,\sigma$, estimated by adding in quadrature the errors of the bins) between the lowest and highest of our redshift bins. This result suggests, that as we move to higher redshifts, galaxies that host AGN tend to have a higher increase of their SFR compared to star-forming galaxies. At higher redshifts, galaxy mergers are more common than at low redshifts \citep[e.g.][]{Lin2008}. These mergers can drive gas towards the centre of galaxies leading to enhanced star-formation and AGN activity when some of this gas is deposited on the central black hole, as has been shown by both theoretical \citep[e.g.][]{Barnes1992, Hopkins2006, Hopkins2008} and observational studies \citep[e.g.][]{Hung2013}. Galaxies in the late stages of a merger have increased SFR and AGN activity \citep[e.g.][]{Stierwalt2013}. Therefore, our findings that SFR of AGN increases more rapidly with redshift compared to non-AGN systems, may indicate that AGN are observed at a late(r) evolutionary merger stage when the central black hole has been activated and the SFR of the galaxy is at its peak. Alternatively, the enhanced SFR of AGN compared to normal galaxies at high redshifts, may be due to the different AGN fueling processes and different star formation trigger mechanisms at high and low redshifts \citep[e.g. mergers vs. disc instabilities][]{Somerville2001, Keres2005}.

\cite{Mullaney2015}, used X-ray AGN from $\it{Chandra}$ Deep Field South and $\it{Chandra}$ Deep Field North and found no evolution of SFR$_{norm}$ with redshift. However, their sample is extracted from narrow fields and therefore does not probe as high and many luminous sources as those detected in XXL. Based on our results, presented in Fig. \ref{comparison_bins}, the SFR$_{norm}$ evolution becomes apparent at high X-ray luminosities  ($\rm L_X \geq 10^{44}$ $\rm erg s^{-1}$), that are rare in deep fields.

\subsection{Star formation as a function of X-ray absorption}
\label{sec_sfr_nh}

SFR is a galaxy wide quantity while X-ray obscuration occurs in the regime around the black hole. However, it has been claimed that obscuration can also occur in galaxy scale \citep[e.g.][]{Fabbiano2017, Malizia2020a}. In the latter case, the two properties may be correlated. Thus, in this part of our analysis, we investigate whether there is a dependence of SFR with the X-ray absorption. 

\cite{Rosario2012} used a sample of AGNs from GOODS-South, GOODS-North and COSMOS fields, spanning the redshifts 0.2 < z < 2.5. The $\rm N_H$ values were estimated using either spectral fits for X-ray sources with sufficient counts or scalings based on hardness ratios for faint X-ray sources. They found a mild dependence between the mean far IR luminosity (SFR proxy) and the X-ray obscuring column, $\rm N_H$. On the other hand, \cite{Rovilos2012}used AGNs from the 3Ms $\it{XMM-Newton}$ survey with z $\simeq$ 0.5$-$4 and reported that there is no correlation between SFR and $\rm N_H$. They claimed that the absorption is likely to be linked to the nuclear region rather than the host galaxy. Their findings were recently confirmed, by \cite{Stemo2020}. This group used X-ray and/or IR selected AGN (Spitzer and Chandra data) and composed a sample of 2585 sources with redshift range 0.2 < z < 2.5. They compiled data come from the GEMS, COSMOS, GOODS and AEGIS surveys. They used extinction parameter, $E_{B-V}$, values (estimated through SED fitting) to infer the $\rm N_H$ values, using a conversion factor of E$_{B-V}$/ $\rm N_H$ = 1.80$\pm$0.15$\times$10$^{-23}$. Based on their findings, the relation between the SFR and $\rm N_H$ is flat,  up to $\rm z=2.5$. According to their interpretation, this behaviour indicates a difference in fuelling processes or timescales between SMBH growth and host galaxy star formation. 

\begin{figure}
\begin{center}  
\includegraphics[width=1.\linewidth]{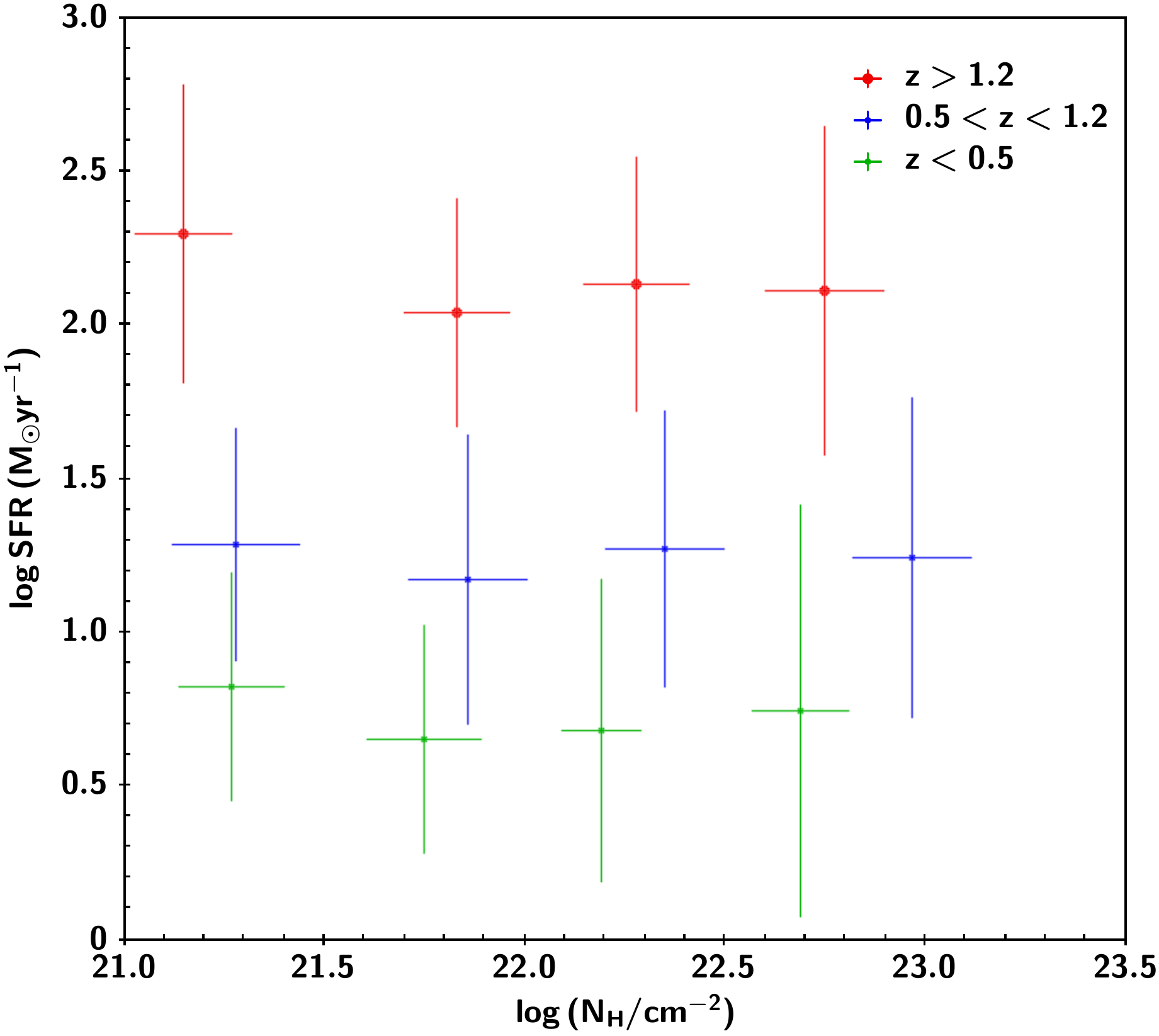}
\end{center}
  \caption{SFR as a function of $\rm N_H$. The sample is split into redshift bins. Green, blue and red colours refer to $\rm z<0.5$, $\rm 0.5<z<1.2$ and $\rm z>1.2$, respectively.  Median SFR and $\rm N_H$ values are presented. The error bars represent the 1$\sigma$ dispersion of each bin. Measurements reveal a flat relation between star-formation and X-ray absorption, at all redshift ranges.}
  \label{frac_SFR}
\end{figure}

In Fig. \ref{frac_SFR}, we plot the star formation as a function of $\rm N_H$ for our X-ray AGN sources. All sources are weighted based on the $\frac{1}{V_{max}}$ method, as well as on the significance of the SFR measurements. Median SFR and $\rm N_H$ values are presented. The error bars represent the 1\,$\sigma$ dispersion of each bin, i.e., they do not take into account the errors of the individual SFR calculations. Measurements are binned in $\rm N_H$, with bin size of 0.5\,dex. We detect a flat relation between the two parameters, at all redshifts spanned by our sample. Our results are in agreement with previous works that have used AGN with similar X-ray properties, i.e. low to moderate levels of X-ray absorption. We note, however, that this result may not hold at higher $\rm N_H$ values.

\section{Summary}
In this study, we use 3,213 AGNs from the $\it{XMM}$-XXL northern field to investigate the relation of the AGN type with the host galaxy properties. About 60\% of our sources have available spectroscopic redshift while for the remaining we use photometric redshifts estimated through a machine learning technique (TPZ). The host galaxy properties, SFR and stellar mass  are estimated via the SED fitting code CIGALE. A statistical method, based on bayesian statistics (BEHR), is applied to derive the HRs for the examined sample. AGN with $\rm N_H > 10^{21.5}\, \rm{cm^{-2}}$ are considered as absorbed. 

First, we study whether there is a connection between  AGN type and the properties of the  host galaxy. We estimate the SFR and M$_*$ distributions of both X-ray absorbed and unabsorbed AGN. A KS-test reveals that the galaxy properties of the two AGN populations are similar. 

Next, we disentangle the effects of M$_*$ and redshift on SFR and examine the SFR-$\rm L_X$ relation, as a function of AGN type. Based on our findings, the SFR-$\rm L_X$ relation is similar for absorbed and unabsorbed AGN, at all redshifts spanned by our sample. This behaviour indicates that the interplay between the AGN and its host galaxy is independent of the obscuration.

Finally, we explore whether the SFR varies as a function of the absorbing column density. Our results show that there is no relation between the two, suggesting that either the two processes take place in different scales or that even if absorption extents to galactic scales does not relate with the SFR of the host galaxy.

Overall, our results suggest that there is no connection between X-ray absorption and the properties of the host galaxy nor the AGN-galaxy co-evolution. This, provides support to the unification model, i.e. X-ray absorption seems to be an inclination effect and not a phase in the lifetime of the AGN. We note, however, that in our analysis we have adopted a rather low X-ray absorption limit. $\it{XMM}$-XXL is a shallow exposure field, and thus there is only a small number  of heavily absorbed sources ($\rm N_H \sim 10^{23}\, \rm{cm^{-2}}$).  Future surveys that will provide larger datasets of obscured AGN and/or with higher column densities (eROSITA, ATHENA) will allow us to study whether this picture holds also for the most  heavily absorbed AGN.

\begin{acknowledgements}
The authors thank the anonymous referee for their detailed report that improved the quality of the paper.
The authors also thank Dr. A. Akylas for his help in estimating the X-ray properties of the sample and Prof. P. Papadopoulos for a detailed reading of the manuscript and useful comments. \\
VAM and IG acknowledge support of this work by the PROTEAS II project (MIS 5002515), which is implemented under the "Reinforcement of the Research and Innovation Infrastructure" action, funded by the "Competitiveness, Entrepreneurship and Innovation" operational programme (NSRF 2014-2020) and co-financed by Greece and the European Union (European Regional Development Fund). \\
This research is co-financed by Greece and the European Union (European Social Fund-ESF) through the Operational Programme "Human Resources Development, Education and Lifelong Learning 2014-2020" in the context of the project "Anatomy of galaxies: their stellar and dust content though cosmic time" (MIS 5052455).
\\
GM acknowledges support by the Agencia Estatal de Investigación, Unidad de Excelencia María de Maeztu, ref. MDM-2017-0765.
\\
$\it{XXL}$  is  an  international  project  based  around  an  $\it{XXM}$  Very Large Programme surveying two 25 deg$^2$ extragalactic fields at a depth of $\sim$  6 $\times$ $10^{-15}$ erg cm $^{-2}$ s$^{-1}$ in the [0.5-2] keV band for point-like sources. The XXL website ishttp://irfu.cea.fr/xxl/.  Multi-band  information  and  spectroscopic  follow-up  of the X-ray sources are obtained through a number of survey programmes,  summarised  at http://xxlmultiwave.pbworks.com/.
\\
This research has made use of data obtained from the 3XMM XMM-\textit{Newton} 
serendipitous source catalogue compiled by the 10 institutes of the XMM-\textit{Newton} 
Survey Science Centre selected by ESA.
\\
This work is based on observations made with XMM-\textit{Newton}, an ESA science mission with instruments and contributions directly funded by ESA Member States 
and NASA. 
\\
Funding for the Sloan Digital Sky Survey IV has been provided by the Alfred P. Sloan Foundation, the U.S. Department of Energy Office of Science, and the Participating Institutions. SDSS-IV acknowledges
support and resources from the Center for High-Performance Computing at
the University of Utah. The SDSS web site is \url{www.sdss.org}.
\\
SDSS-IV is managed by the Astrophysical Research Consortium for the 
Participating Institutions of the SDSS Collaboration including the 
Brazilian Participation Group, the Carnegie Institution for Science, 
Carnegie Mellon University, the Chilean Participation Group, the French Participation Group, Harvard-Smithsonian Center for Astrophysics, 
Instituto de Astrof\'isica de Canarias, The Johns Hopkins University, 
Kavli Institute for the Physics and Mathematics of the Universe (IPMU) / 
University of Tokyo, Lawrence Berkeley National Laboratory, 
Leibniz Institut f\"ur Astrophysik Potsdam (AIP),  
Max-Planck-Institut f\"ur Astronomie (MPIA Heidelberg), 
Max-Planck-Institut f\"ur Astrophysik (MPA Garching), 
Max-Planck-Institut f\"ur Extraterrestrische Physik (MPE), 
National Astronomical Observatories of China, New Mexico State University, 
New York University, University of Notre Dame, 
Observat\'ario Nacional / MCTI, The Ohio State University, 
Pennsylvania State University, Shanghai Astronomical Observatory, 
United Kingdom Participation Group,
Universidad Nacional Aut\'onoma de M\'exico, University of Arizona, 
University of Colorado Boulder, University of Oxford, University of Portsmouth, 
University of Utah, University of Virginia, University of Washington, University of Wisconsin, 
Vanderbilt University, and Yale University.
\\
This publication makes use of data products from the Wide-field Infrared Survey 
Explorer, which is a joint project of the University of California, Los Angeles, 
and the Jet Propulsion Laboratory/California Institute of Technology, funded by 
the National Aeronautics and Space Administration. 
\\
The VISTA Data Flow System pipeline processing and science archive are described 
in \cite{Irwin2004}, \cite{Hambly2008} and \cite{Cross2012a}. Based on 
observations obtained as part of the VISTA Hemisphere Survey, ESO Program, 
179.A-2010 (PI: McMahon). We have used data from the 3rd data release.
\end{acknowledgements}

\bibliography{mybib}{}

\begin{thebibliography}{105}
\expandafter\ifx\csname natexlab\endcsname\relax\def\natexlab#1{#1}\fi

\bibitem[{Aird {et~al.}(2016)Aird, Coil, \& Georgakakis}]{Aird2016}
Aird, J., Coil, A.~L., \& Georgakakis, A. 2016, MNRAS, 465, 3390

\bibitem[{Aird {et~al.}(2019)Aird, Coil, \& Georgakakis}]{Aird2019}
Aird, J., Coil, A.~L., \& Georgakakis, A. 2019, MNRAS, 484, 4360

\bibitem[{Aird {et~al.}(2015)Aird, Coil, Georgakakis, Nandra, Barro, \&
  P{\'{e}}rez-Gonz{\'{a}}lez}]{Aird2015}
Aird, J., Coil, A.~L., Georgakakis, A., {et~al.} 2015, MNRAS, 451, 1892

\bibitem[{{Akylas} \& {Georgantopoulos}(2008)}]{Akylas2008}
{Akylas}, A. \& {Georgantopoulos}, I. 2008, A\&A, 479, 735

\bibitem[{{Akylas} {et~al.}(2006){Akylas}, {Georgantopoulos}, {Georgakakis},
  {Kitsionas}, \& {Hatziminaoglou}}]{Akylas2006}
{Akylas}, A., {Georgantopoulos}, I., {Georgakakis}, A., {Kitsionas}, S., \&
  {Hatziminaoglou}, E. 2006, A\&A, 459, 693

\bibitem[{{Alexander} \& {Hickox}(2012)}]{Alexander2012}
{Alexander}, D.~M. \& {Hickox}, R.~C. 2012, NewAR, 56, 93

\bibitem[{Antonucci(1993)}]{Antonucci1993}
Antonucci, R. 1993, ARA\&A, 31, 473

\bibitem[{Barnes \& Hernquist(1992)}]{Barnes1992}
Barnes, J.~E. \& Hernquist, L. 1992, Annual Review of Astronomy and
  Astrophysics, 30, 705

\bibitem[{Barvainis(1987)}]{Barvainis1987}
Barvainis, R. 1987, ApJ, 320, 537

\bibitem[{Bernhard {et~al.}(2019)Bernhard, Grimmett, Mullaney, Daddi,
  Tadhunter, \& Jin}]{Bernhard2018}
Bernhard, E., Grimmett, L.~P., Mullaney, J.~R., {et~al.} 2019, MNRAS, 483, L52

\bibitem[{Boquien {et~al.}(2019)Boquien, Burgarella, Roehlly, Buat, Ciesla,
  Corre, Inoue, \& Salas}]{Boquien2019}
Boquien, M., Burgarella, D., Roehlly, Y., {et~al.} 2019, A{\&}A, 622, A103

\bibitem[{Bournaud {et~al.}(2007)Bournaud, Elmegreen, \&
  Elmegreen}]{Bournaud2007}
Bournaud, F., Elmegreen, B.~G., \& Elmegreen, D.~M. 2007, ApJ, 670, 237

\bibitem[{Brinchmann {et~al.}(2004)Brinchmann, Charlot, White, Tremonti,
  Kauffmann, Heckman, \& Brinkmann}]{Brinchmann2004}
Brinchmann, J., Charlot, S., White, S. D.~M., {et~al.} 2004, MNRAS, 351, 1151

\bibitem[{Brown {et~al.}(2019)Brown, Nayyeri, Cooray, Ma, Hickox, \&
  Azadi}]{Brown2019}
Brown, A., Nayyeri, H., Cooray, A., {et~al.} 2019, The Astrophysical Journal,
  871, 87

\bibitem[{Buchner \& Bauer(2016)}]{Buchner2016}
Buchner, J. \& Bauer, F.~E. 2016, MNRAS, 465, 4348

\bibitem[{Burgarella {et~al.}(2005)Burgarella, Buat, \&
  Iglesias-P{\'{a}}ramo}]{Burgarella2005}
Burgarella, D., Buat, V., \& Iglesias-P{\'{a}}ramo, J. 2005, MNRAS, 360, 1413

\bibitem[{Carrasco~Kind \& {Brunner}(2013)}]{Kind2013}
Carrasco~Kind, M. \& {Brunner}, R.~J. 2013, MNRAS, 432, 1483

\bibitem[{Chen {et~al.}(2013)Chen, Hickox, Alberts, Brodwin, Jones, Murray,
  Alexander, Assef, Brown, Dey, Forman, Gorjian, Goulding, Floch, Jannuzi,
  Mullaney, \& Pope}]{Chen2013}
Chen, C.-T.~J., Hickox, R.~C., Alberts, S., {et~al.} 2013, ApJ, 773, 3

\bibitem[{Chen {et~al.}(2015)Chen, Hickox, Alberts, Harrison, Alexander, Assef,
  Brodwin, Brown, Moro, Forman, Gorjian, Goulding, Hainline, Jones, Kochanek,
  Murray, Pope, Rovilos, \& Stern}]{Chen2015}
Chen, C.-T.~J., Hickox, R.~C., Alberts, S., {et~al.} 2015, ApJ, 802, 50

\bibitem[{Ciesla {et~al.}(2015)Ciesla, Charmandaris, Georgakakis, Bernhard,
  Mitchell, Buat, Elbaz, LeFloc'h, Lacey, Magdis, \& Xilouris}]{Ciesla2015}
Ciesla, L., Charmandaris, V., Georgakakis, A., {et~al.} 2015, A\&A, 576, A10

\bibitem[{Ciotti \& Ostriker(1997)}]{Ciotti1997}
Ciotti, L. \& Ostriker, J.~P. 1997, ApJ, 487, L105

\bibitem[{Ciotti \& Ostriker(2001)}]{Ciotti2001}
Ciotti, L. \& Ostriker, J.~P. 2001, ApJ, 551, 131

\bibitem[{Circosta {et~al.}(2019)Circosta, Vignali, Gilli, Feltre, Vito,
  Calura, Mainieri, Massardi, \& Norman}]{Circosta2019}
Circosta, C., Vignali, C., Gilli, R., {et~al.} 2019, A\&A, 623, A172

\bibitem[{Civano {et~al.}(2016)Civano, Marchesi, Comastri, Urry, Elvis,
  Cappelluti, Puccetti, Brusa, Zamorani, Hasinger, Aldcroft, Alexander,
  Allevato, Brunner, Capak, Finoguenov, Fiore, Fruscione, Gilli, Glotfelty,
  Griffiths, Hao, Harrison, Jahnke, Kartaltepe, Karim, LaMassa, Lanzuisi,
  Miyaji, Ranalli, Salvato, Sargent, Scoville, Schawinski, Schinnerer,
  Silverman, Smolcic, Stern, Toft, Trakhenbrot, Treister, \&
  Vignali}]{Civano2016}
Civano, F., Marchesi, S., Comastri, A., {et~al.} 2016, ApJ, 819, 62

\bibitem[{Cross {et~al.}(2012)Cross, Collins, Mann, Read, Sutorius, Blake,
  Holliman, Hambly, Emerson, Lawrence, \& Noddle}]{Cross2012a}
Cross, N. J.~G., Collins, R.~S., Mann, R.~G., {et~al.} 2012, A\&A, 548, A119

\bibitem[{Daddi {et~al.}(2007)Daddi, Dickinson, Morrison, Chary, Cimatti,
  Elbaz, Frayer, Renzini, Pope, Alexander, Bauer, Giavalisco, Huynh, Kurk, \&
  Mignoli}]{Daddi2007}
Daddi, E., Dickinson, M., Morrison, G., {et~al.} 2007, ApJ, 670, 156

\bibitem[{{Di Matteo} {et~al.}(2008){Di Matteo}, {Colberg}, {Springel},
  {Hernquist}, \& {Sijacki}}]{DiMatteo2008}
{Di Matteo}, T., {Colberg}, J., {Springel}, V., {Hernquist}, L., \& {Sijacki},
  D. 2008, ApJ, 676, 33

\bibitem[{{Di Matteo} {et~al.}(2005){Di Matteo}, {Springel}, \&
  {Hernquist}}]{DiMatteo2005}
{Di Matteo}, T., {Springel}, V., \& {Hernquist}, L. 2005, Nature, 433, 604

\bibitem[{Draine(2003)}]{Draine2003}
Draine, B.~T. 2003, The Astrophysical Journal, 598, 1026

\bibitem[{Elbaz {et~al.}(2007)Elbaz, Daddi, Borgne, Dickinson, Alexander,
  Chary, Starck, Brandt, Kitzbichler, MacDonald, Nonino, Popesso, Stern, \&
  Vanzella}]{Elbaz2007}
Elbaz, D., Daddi, E., Borgne, D.~L., {et~al.} 2007, A{\&}A, 468, 33

\bibitem[{Elvis {et~al.}(1994)Elvis, Wilkes, McDowell, Green, Bechtold,
  Willner, Oey, Polomski, \& Cutri}]{Elvis1994}
Elvis, M., Wilkes, B.~J., McDowell, J.~C., {et~al.} 1994, ApJS, 95, 1

\bibitem[{Fabbiano {et~al.}(2017)Fabbiano, Elvis, Paggi, Karovska, Maksym,
  Raymond, Risaliti, \& Wang}]{Fabbiano2017}
Fabbiano, G., Elvis, M., Paggi, A., {et~al.} 2017, ApJ, 842, L4

\bibitem[{Fabian(1999)}]{Fabian1999}
Fabian, A.~C. 1999, MNRAS, 308, L39

\bibitem[{{Fanidakis} {et~al.}(2011){Fanidakis}, {Baugh}, {Benson}, {Bower},
  {Cole}, {Done}, \& {Frenk}}]{Fanidakis2011a}
{Fanidakis}, N., {Baugh}, C.~M., {Benson}, A.~J., {et~al.} 2011, MNRAS, 410, 53

\bibitem[{{Ferrarese} \& {Merritt}(2000)}]{Ferrarese2000}
{Ferrarese}, L. \& {Merritt}, D. 2000, ApJ, 539, 9

\bibitem[{Florez {et~al.}(2020)Florez, Jogee, Sherman, Stevans, Finkelstein,
  Papovich, Kawinwanichakij, Ciardullo, Gronwall, Urry, Kirkpatrick, LaMassa,
  Ananna, \& Wold}]{Florez2020}
Florez, J., Jogee, S., Sherman, S., {et~al.} 2020, Monthly Notices of the Royal
  Astronomical Society, 497, 3273

\bibitem[{{Fritz} {et~al.}(2006){Fritz}, {Franceschini}, \&
  {Hatziminaoglou}}]{Fritz2006}
{Fritz}, J., {Franceschini}, A., \& {Hatziminaoglou}, E. 2006, MNRAS, 166, 767

\bibitem[{Georgakakis \& Nandra(2011)}]{Georgakakis2011a}
Georgakakis, A. \& Nandra, K. 2011, Monthly Notices of the Royal Astronomical
  Society, 414, 992

\bibitem[{Georgakakis {et~al.}(2017)Georgakakis, Salvato, Liu, Buchner, Brandt,
  Ananna, Schulze, Shen, LaMassa, Nandra, Merloni, \&
  McGreer}]{Georgakakis2017}
Georgakakis, A., Salvato, M., Liu, Z., {et~al.} 2017, MNRAS, 469, 3232

\bibitem[{{Gilli} {et~al.}(2007){Gilli}, {Comastri}, \& {Hasinger}}]{Gilli2007}
{Gilli}, R., {Comastri}, A., \& {Hasinger}, G. 2007, A\&A, 463, 79

\bibitem[{Grimmett {et~al.}(2020)Grimmett, Mullaney, Bernhard, Harrison,
  Alexander, Stanley, Masoura, \& Walters}]{Grimmett2020}
Grimmett, L.~P., Mullaney, J.~R., Bernhard, E.~P., {et~al.} 2020, MNRAS, 495,
  1392

\bibitem[{Hambly {et~al.}(2008)Hambly, Collins, Cross, Mann, Read, Sutorius,
  Bond, Bryant, Emerson, Lawrence, Rimoldini, Stewart, Williams, Adamson,
  Hirst, Dye, \& Warren}]{Hambly2008}
Hambly, N.~C., Collins, R.~S., Cross, N. J.~G., {et~al.} 2008, MNRAS, 384, 637

\bibitem[{{Harrison} {et~al.}(2012)}]{Harrison2012}
{Harrison}, C.~M. {et~al.} 2012, ApJL, 760, 5

\bibitem[{Hickox \& Alexander(2018)}]{Hickox2018a}
Hickox, R.~C. \& Alexander, D.~M. 2018, ARA\&A, 56, 625

\bibitem[{Hickox {et~al.}(2007)Hickox, Jones, Forman, Murray, Brodwin, Brown,
  Eisenhardt, Stern, Kochanek, Eisenstein, Cool, Jannuzi, Dey, Brand, Gorjian,
  \& Caldwell}]{Hickox2007}
Hickox, R.~C., Jones, C., Forman, W.~R., {et~al.} 2007, ApJ, 671, 1365

\bibitem[{Hickox {et~al.}(2014)Hickox, Mullaney, Alexander, Chen, Civano, \&
  Goulding}]{Hickox2014}
Hickox, R.~C., Mullaney, J.~R., Alexander, D.~M., {et~al.} 2014, ApJ, 782, 11

\bibitem[{Hickox {et~al.}(2011)Hickox, Myers, Brodwin, Alexander, Forman,
  Jones, Murray, Brown, Cool, Kochanek, Dey, Jannuzi, Eisenstein, Assef,
  Eisenhardt, Gorjian, Stern, Floch, Caldwell, Goulding, \&
  Mullaney}]{Hickox2011}
Hickox, R.~C., Myers, A.~D., Brodwin, M., {et~al.} 2011, ApJ, 731, 117

\bibitem[{Hopkins \& Hernquist(2006)}]{Hopkins2006}
Hopkins, P.~F. \& Hernquist, L. 2006, ApJS, 166, 1

\bibitem[{Hopkins {et~al.}(2008)Hopkins, Hernquist, Cox, \&
  Kere{\v{s}}}]{Hopkins2008}
Hopkins, P.~F., Hernquist, L., Cox, T.~J., \& Kere{\v{s}}, D. 2008, ApJS, 175,
  356

\bibitem[{{Hopkins} {et~al.}(2006){Hopkins}, {Hernquist}, {Cox}, {Robertson},
  {Di Matteo}, \& {Springel}}]{Hopkins2006a}
{Hopkins}, P.~F., {Hernquist}, L., {Cox}, T.~J., {et~al.} 2006, ApJ, 639, 700

\bibitem[{Hung {et~al.}(2013)Hung, Sanders, Casey, Lee, Barnes, Capak,
  Kartaltepe, Koss, Larson, Floch, Lockhart, Man, Mann, Riguccini, Scoville, \&
  Symeonidis}]{Hung2013}
Hung, C.-L., Sanders, D.~B., Casey, C.~M., {et~al.} 2013, The Astrophysical
  Journal, 778, 129

\bibitem[{Hurley {et~al.}(2017)Hurley, Oliver, Betancourt, Clarke, Cowley,
  Duivenvoorden, Farrah, Griffin, Lacey, Floch, Papadopoulos, Sargent, Scudder,
  Vaccari, Valtchanov, \& Wang}]{Hurley2017}
Hurley, P.~D., Oliver, S., Betancourt, M., {et~al.} 2017, Monthly Notices of
  the Royal Astronomical Society, 464, 885

\bibitem[{Irwin {et~al.}(2004)Irwin, Lewis, Hodgkin, Bunclark, Evans, McMahon,
  Emerson, Stewart, \& Beard}]{Irwin2004}
Irwin, M.~J., Lewis, J., Hodgkin, S., {et~al.} 2004, in SPIE

\bibitem[{Kere\v{s} {et~al.}(2005)Kere\v{s}, Katz, Weinberg, \&
  Dave}]{Keres2005}
Kere\v{s}, D., Katz, N., Weinberg, D.~H., \& Dave, R. 2005, Monthly Notices of
  the Royal Astronomical Society, 363, 2

\bibitem[{{Lanzuisi} {et~al.}(2017)}]{Lanzuisi2017}
{Lanzuisi}, G. {et~al.} 2017, A\&A, 602, 13

\bibitem[{Lin {et~al.}(2008)Lin, Patton, Koo, Casteels, Conselice, Faber, Lotz,
  Willmer, Hsieh, Chiueh, Newman, Novak, Weiner, \& Cooper}]{Lin2008}
Lin, L., Patton, D.~R., Koo, D.~C., {et~al.} 2008, The Astrophysical Journal,
  681, 232

\bibitem[{Liu {et~al.}(2016)Liu, Merloni, Georgakakis, Menzel, Buchner, Nandra,
  Salvato, Shen, Brusa, \& Streblyanska}]{Liu2016}
Liu, Z., Merloni, A., Georgakakis, A., {et~al.} 2016, MNRAS, 459, 1602

\bibitem[{Lusso \& Risaliti(2016)}]{Lusso2016}
Lusso, E. \& Risaliti, G. 2016, ApJ, 819, 154

\bibitem[{{Lutz} {et~al.}(2010)}]{Lutz2010}
{Lutz}, D. {et~al.} 2010, ApJ, 712, 1287

\bibitem[{Magdis {et~al.}(2010)Magdis, Rigopoulou, Huang, \&
  Fazio}]{Magdis2010}
Magdis, G.~E., Rigopoulou, D., Huang, J.-S., \& Fazio, G.~G. 2010, MNRAS, 401,
  1521

\bibitem[{Magorrian {et~al.}(1998)Magorrian, Tremaine, Richstone, Bender,
  Bower, Dressler, Faber, Gebhardt, Green, Grillmair, Kormendy, \&
  Lauer}]{Magorrian1998}
Magorrian, J., Tremaine, S., Richstone, D., {et~al.} 1998, AJ, 115, 2285

\bibitem[{Maiolino \& Rieke(1995)}]{Maiolino1995}
Maiolino, R. \& Rieke, G.~H. 1995, ApJ, 454, 95

\bibitem[{Maiolino {et~al.}(1997)Maiolino, Ruiz, Rieke, \&
  Papadopoulos}]{Maiolino1997}
Maiolino, R., Ruiz, M., Rieke, G.~H., \& Papadopoulos, P. 1997, ApJ, 485, 552

\bibitem[{Malizia {et~al.}(2020)Malizia, Bassani, Stephen, Bazzano, \&
  Ubertini}]{Malizia2020a}
Malizia, A., Bassani, L., Stephen, J.~B., Bazzano, A., \& Ubertini, P. 2020,
  A{\&}A, 639, A5

\bibitem[{Malkan {et~al.}(1998)Malkan, Gorjian, \& Tam}]{Malkan1998}
Malkan, M.~A., Gorjian, V., \& Tam, R. 1998, Astrophys. J., Suppl. Ser., 117,
  25

\bibitem[{Masoura {et~al.}(2020)Masoura, Georgantopoulos, Mountrichas, Vignali,
  Koulouridis, Chiappetti, Fotopoulou, Paltani, \& Pierre}]{Masoura2020}
Masoura, V.~A., Georgantopoulos, I., Mountrichas, G., {et~al.} 2020, Astronomy
  {\&} Astrophysics, 638, A45

\bibitem[{Masoura {et~al.}(2018)Masoura, Mountrichas, Georgantopoulos, Ruiz,
  Magdis, \& Plionis}]{Masoura2018}
Masoura, V.~A., Mountrichas, G., Georgantopoulos, I., {et~al.} 2018, A\&A, 618,
  A31

\bibitem[{Matt(2000)}]{Matt2000}
Matt, G. 2000, A\&A, 355, L31

\bibitem[{Mendez {et~al.}(2016)Mendez, Coil, Aird, Skibba, Diamond-Stanic,
  Moustakas, Blanton, Cool, Eisenstein, Wong, \& Zhu}]{Mendez2016}
Mendez, A.~J., Coil, A.~L., Aird, J., {et~al.} 2016, ApJ, 821, 55

\bibitem[{Menzel {et~al.}(2016)Menzel, Merloni, Georgakakis, Salvato, Aubourg,
  Brandt, Brusa, Buchner, Dwelly, Nandra, P{\^a}ris, Petitjean, \&
  Schwope}]{Menzel2016}
Menzel, M.-L., Merloni, A., Georgakakis, A., {et~al.} 2016, MNRAS, 457, 110

\bibitem[{Merloni {et~al.}(2014)Merloni, Bongiorno, Brusa, Iwasawa, Mainieri,
  Magnelli, Salvato, Berta, Cappelluti, Comastri, Fiore, Gilli, Koekemoer,
  Floc, Lusso, Lutz, Miyaji, Pozzi, Riguccini, Rosario, Silverman, Symeonidis,
  Treister, Vignali, \& Zamorani}]{Merloni2014}
Merloni, A., Bongiorno, A., Brusa, M., {et~al.} 2014, MNRAS, 437, 3550

\bibitem[{Mountrichas {et~al.}(2017)Mountrichas, Georgantopoulos, Secrest,
  Ordov{\'{a}}s-Pascual, Corral, Akylas, Mateos, Carrera, \&
  Batziou}]{Mountrichas2017}
Mountrichas, G., Georgantopoulos, I., Secrest, N.~J., {et~al.} 2017, MNRAS,
  468, 3042

\bibitem[{{Mountrichas} {et~al.}(2016)}]{Mountrichas2016}
{Mountrichas}, G. {et~al.} 2016, MNRAS, 457, 4195

\bibitem[{Mukai(1993)}]{Mukai1993}
Mukai, K. 1993, Legacy, 3, 21

\bibitem[{Mullaney {et~al.}(2015)Mullaney, Alexander, Aird, Bernhard, Daddi,
  Moro, Dickinson, Elbaz, Harrison, Juneau, Liu, Pannella, Rosario, Santini,
  Sargent, Schreiber, Simpson, \& Stanley}]{Mullaney2015}
Mullaney, J.~R., Alexander, D.~M., Aird, J., {et~al.} 2015, MNRAS, 453, L83

\bibitem[{Netzer(2015)}]{Netzer2015}
Netzer, H. 2015, ARA\&A, 53, 365

\bibitem[{Noll {et~al.}(2009)Noll, Burgarella, Giovannoli, Buat, Marcillac, \&
  Mu{\~{n}}oz-Mateos}]{Noll2009}
Noll, S., Burgarella, D., Giovannoli, E., {et~al.} 2009, A\&A, 507, 1793

\bibitem[{Oliver {et~al.}(2012)Oliver, Bock, Altieri, Amblard, Arumugam,
  Aussel, Babbedge, Beelen, Bethermin, Blain, Boselli, Bridge, Brisbin, Buat,
  Burgarella, Castro-Rodriguez, Cava, Chanial, Cirasuolo, Clements, Conley,
  Conversi, Cooray, Dowell, Dubois, Dwek, Dye, Eales, Elbaz, Farrah, Feltre,
  Ferrero, Fiolet, Fox, Franceschini, Gear, Giovannoli, Glenn, Gong, Solares,
  Griffin, Halpern, Harwit, Hatziminaoglou, Heinis, Hurley, Hwang, Hyde, Ibar,
  Ilbert, Isaak, Ivison, Lagache, Floch, Levenson, Faro, Lu, Madden, Maffei,
  Magdis, Mainetti, Marchetti, Marsden, Marshall, Mortier, Nguyen, Halloran,
  Omont, Page, Panuzzo, Papageorgiou, Patel, Pearson, P{\'{e}}rez-Fournon,
  Pohlen, Rawlings, Raymond, Rigopoulou, Riguccini, Rizzo, Rodighiero,
  Roseboom, Rowan-Robinson, Portal, Schulz, Scott, Seymour, Shupe, Smith,
  Stevens, Symeonidis, Trichas, Tugwell, Vaccari, Valtchanov, Vieira, Viero,
  Vigroux, Wang, Ward, Wardlow, Wright, Xu, \& Zemcov}]{Oliver2012a}
Oliver, S.~J., Bock, J., Altieri, B., {et~al.} 2012, Monthly Notices of the
  Royal Astronomical Society, 424, 1614

\bibitem[{{Oliver} {et~al.}(2012)}]{Oliver2012}
{Oliver}, S.~J. {et~al.} 2012, MNRAS, 424, 1614

\bibitem[{Page \& Carrera(2000)}]{Page2000}
Page, M.~J. \& Carrera, F.~J. 2000, MNRAS, 311, 433

\bibitem[{{Page} {et~al.}(2012)}]{Page2012}
{Page}, M.~J. {et~al.} 2012, nat, 485, 213

\bibitem[{Park {et~al.}(2006)Park, Kashyap, Siemiginowska, van Dyk, Zezas,
  Heinke, \& Wargelin}]{Park2006}
Park, T., Kashyap, V.~L., Siemiginowska, A., {et~al.} 2006, AJ, 652, 610

\bibitem[{Pierre {et~al.}(2016)Pierre, Pacaud, Adami, Alis, Altieri, Baran,
  Benoist, Birkinshaw, Bongiorno, Bremer, Brusa, Butler, Ciliegi, Chiappetti,
  Clerc, Corasaniti, Coupon, Breuck, Democles, Desai, Delhaize, Devriendt,
  Dubois, Eckert, Elyiv, Ettori, Evrard, Faccioli, Farahi, Ferrari, Finet,
  Fotopoulou, Fourmanoit, Gandhi, Gastaldello, Gastaud, Georgantopoulos, Giles,
  Guennou, Guglielmo, Horellou, Husband, Huynh, Iovino, Kilbinger, Koulouridis,
  Lavoie, Brun, Fevre, Lidman, Lieu, Lin, Mantz, Maughan, Maurogordato,
  McCarthy, McGee, Melin, Melnyk, Menanteau, Novak, Paltani, Plionis,
  Poggianti, Pomarede, Pompei, Ponman, Ramos-Ceja, Ranalli, Rapetti,
  Raychaudury, Reiprich, Rottgering, Rozo, Rykoff, Sadibekova, Santos,
  Sauvageot, Schimd, Sereno, Smith, Smol{\v{c}}i{\'{c}}, Snowden, Spergel,
  Stanford, Surdej, Valageas, Valotti, Valtchanov, Vignali, Willis, \&
  Ziparo}]{Pierre2016}
Pierre, M., Pacaud, F., Adami, C., {et~al.} 2016, A\&A, 592, A1

\bibitem[{Rodighiero {et~al.}(2015)Rodighiero, Brusa, Daddi, Negrello,
  Mullaney, Delvecchio, Lutz, Renzini, Franceschini, Baronchelli, Pozzi,
  Gruppioni, Strazzullo, Cimatti, \& Silverman}]{Rodighiero2015}
Rodighiero, G., Brusa, M., Daddi, E., {et~al.} 2015, ApJ, 800, L10

\bibitem[{Rosario {et~al.}(2013)Rosario, Trakhtenbrot, Lutz, Netzer, Trump,
  Silverman, Schramm, Lusso, Berta, Bongiorno, Brusa, {et~al.}}]{Rosario2013}
Rosario, D.~J., Trakhtenbrot, B., Lutz, D., {et~al.} 2013, A\&A, 560, A72

\bibitem[{Rosario {et~al.}(2012)}]{Rosario2012}
Rosario, D.~J. {et~al.} 2012, A\&A, 545, 18

\bibitem[{{Rovilos} \& {Georgantopoulos}(2007)}]{Rovilos2007}
{Rovilos}, E. \& {Georgantopoulos}, I. 2007, A\&A, 475, 115

\bibitem[{{Rovilos} {et~al.}(2012)}]{Rovilos2012}
{Rovilos}, E. {et~al.} 2012, A\&A, 546, 16

\bibitem[{Ruiz {et~al.}(2018)Ruiz, Corral, Mountrichas, \&
  Georgantopoulos}]{Ruiz2018}
Ruiz, A., Corral, A., Mountrichas, G., \& Georgantopoulos, I. 2018, A{\&}A,
  618, A52

\bibitem[{Salmon {et~al.}(2015)Salmon, Papovich, Finkelstein, Tilvi, Finlator,
  Behroozi, Dahlen, Dav{\'{e}}, Dekel, Dickinson, Ferguson, Giavalisco, Long,
  Lu, Mobasher, Reddy, Somerville, \& Wechsler}]{Salmon2015}
Salmon, B., Papovich, C., Finkelstein, S.~L., {et~al.} 2015, ApJ, 799, 183

\bibitem[{Sanders {et~al.}(1989)Sanders, Phinney, Neugebauer, Soifer, \&
  Matthews}]{Sanders1989}
Sanders, D.~B., Phinney, E.~S., Neugebauer, G., Soifer, B.~T., \& Matthews, K.
  1989, ApJ, 347, 29

\bibitem[{Schmidt(1968)}]{Schmidt1968}
Schmidt, M. 1968, ApJ, 151, 393

\bibitem[{Schreiber {et~al.}(2015)}]{Schreiber2015}
Schreiber, C. {et~al.} 2015, A\&A, 575, 29

\bibitem[{{Somerville} {et~al.}(2008){Somerville}, {Hopkins}, J., {Robertson},
  \& L.}]{Somerville2008}
{Somerville}, R.~S., {Hopkins}, P.~F., J., C.~T., {Robertson}, B.~E., \& L., H.
  2008, MNRAS, 391, 481

\bibitem[{Somerville {et~al.}(2001)Somerville, Primack, \&
  Faber}]{Somerville2001}
Somerville, R.~S., Primack, J.~R., \& Faber, S.~M. 2001, Monthly Notices of the
  Royal Astronomical Society, 320, 504

\bibitem[{Stanley {et~al.}(2015)Stanley, Harrison, Alexander, Swinbank, Aird,
  Moro, Hickox, \& Mullaney}]{Stanley2015}
Stanley, F., Harrison, C.~M., Alexander, D.~M., {et~al.} 2015, MNRAS, 453, 591

\bibitem[{Stemo {et~al.}(2020)Stemo, Comerford, Barrows, Stern, Assef, \&
  Griffith}]{Stemo2020}
Stemo, A., Comerford, J.~M., Barrows, R.~S., {et~al.} 2020, ApJ, 888, 78

\bibitem[{Stierwalt {et~al.}(2013)Stierwalt, Armus, Surace, Inami, Petric,
  Diaz-Santos, Haan, Charmandaris, Howell, Kim, Marshall, Mazzarella, Spoon,
  Veilleux, Evans, Sanders, Appleton, Bothun, Bridge, Chan, Frayer, Iwasawa,
  Kewley, Lord, Madore, Melbourne, Murphy, Rich, Schulz, Sturm, U, Vavilkin, \&
  Xu}]{Stierwalt2013}
Stierwalt, S., Armus, L., Surace, J.~A., {et~al.} 2013, The Astrophysical
  Journal Supplement Series, 206, 1

\bibitem[{{Sutherland}(1992)}]{Sutherland_and_Saunders1992}
{Sutherland}, W. \&~{Saunders}, W. 1992, MNRAS, 259, 413

\bibitem[{Treister {et~al.}(2004)Treister, Urry, Chatzichristou, Bauer,
  Alexander, Koekemoer, Duyne, Brandt, Bergeron, Stern, Moustakas, Chary,
  Conselice, Cristiani, \& Grogin}]{Treister2004}
Treister, E., Urry, C.~M., Chatzichristou, E., {et~al.} 2004, ApJ, 616, 123

\bibitem[{Treister {et~al.}(2009)Treister, Urry, \& Virani}]{Treister2009}
Treister, E., Urry, C.~M., \& Virani, S. 2009, ApJ, 696, 110

\bibitem[{Urry \& Padovani(1995)}]{Urry1995}
Urry, C.~M. \& Padovani, P. 1995, PASP, 107, 803

\bibitem[{Volonteri {et~al.}(2015{\natexlab{a}})Volonteri, Capelo, Netzer,
  Bellovary, Dotti, \& Governato}]{Volonteri2015}
Volonteri, M., Capelo, P.~R., Netzer, H., {et~al.} 2015{\natexlab{a}}, MNRAS,
  452, L6

\bibitem[{Volonteri {et~al.}(2015{\natexlab{b}})Volonteri, Capelo, Netzer,
  Bellovary, Dotti, \& Governato}]{Volonteri2015a}
Volonteri, M., Capelo, P.~R., Netzer, H., {et~al.} 2015{\natexlab{b}}, MNRAS,
  449, 1470

\bibitem[{Zou {et~al.}(2019)Zou, Yang, Brandt, \& Xue}]{Zou2019}
Zou, F., Yang, G., Brandt, W.~N., \& Xue, Y. 2019, ApJ, 878, 11

\end{thebibliography}
\bibliographystyle{aa}

\end{document}